\documentclass[12pt,a4paper]{amsart}

\usepackage{empheq}
\usepackage{tcolorbox}
\usepackage{framed}
\usepackage{amsmath}
\usepackage[latin9]{inputenc}
\usepackage{amstext}
\usepackage{amssymb}
\usepackage{caption}
\usepackage{longtable}
\usepackage{lscape}
\usepackage{tabularx}
\usepackage{varioref}
\usepackage{amsthm}
\usepackage{graphicx}
\usepackage{txfonts}
\usepackage{pxfonts}
\usepackage{marginnote}
\usepackage{epic}
\usepackage{eepic}
\usepackage{float}
\usepackage{rotating}
\usepackage{epsfig}
\usepackage{indentfirst}
\usepackage{array}
\usepackage{varioref}
\usepackage{tikz}
\usepackage{marginnote}
\usepackage{pgf}
\usepackage{bbm}
\usepackage{appendix}
\usepackage{amsbsy}
\usepackage{latexsym}
\usepackage{amsfonts}
\usepackage{pstricks}
\usepackage{color}
\usepackage[numbers,square,comma, compress]{natbib}
\usepackage{eurosym}
\usepackage{tikz}
\usepackage{pdfpages}
\usepackage{wasysym}
\usepackage{subcaption}
\usepackage{mathdots}

\newcommand{\ws}{\square}

\newcommand{\eo}{\epsilon_1}
\newcommand{\et}{\epsilon_2}

\usepackage[scale=0.81]{geometry}

\makeatletter

\pdfpageheight\paperheight
\pdfpagewidth\paperwidth

\numberwithin{equation}{section}
\numberwithin{figure}{section}

\numberwithin{equation}{section}
\numberwithin{figure}{section}

\let\emptyset\varnothing

\def\ll{\left\lgroup}
\def\rr{\right\rgroup}

\def\ll{ \left\lgroup}
\def\rr{\right\rgroup}

\newcommand{\cC}{\mathcal{C}}

\newcommand{\cM}{\mathcal{M}}
\newcommand{\cN}{\mathcal{N}}

\newcommand{\cR}{\mathcal{R}}

\newcommand{\CC}{\mathbbm{C}}

\newcommand{\RR}{\mathbbm{R}}
\newcommand{\ZZ}{\mathbbm{Z}}

\newcommand{\1}{{\bf 1.}}
\newcommand{\2}{{\bf 2.}}
\newcommand{\3}{{\bf 3.}}
\newcommand{\4}{{\bf 4.}}
\newcommand{\5}{{\bf 5.}}

\newcommand{\wsq}{\square}

\newcommand{\bea}{\begin{eqnarray}\displaystyle}
\newcommand{\eea}{\end{eqnarray}}

\newcommand{\figref}[1]{Fig.~\protect\ref{#1}}

\def\bi{\pmb{\iota }}

\restylefloat{figure}

\begin{document} 
	
\title[]{Macdonald topological vertices and brane condensates}
	
\author[]{Omar Foda          \!$^{{\scriptstyle {\, 1            }}}$ and
          Masahide Manabe    \!$^{{\scriptstyle {\, 2            }}}$
          }
		  
\address{
\!\!\!\!\!\!\!\!\!
$^{{\scriptstyle 1}}$
Mathematics and Statistics, University of Melbourne,
Royal Parade, Parkville, VIC 3010, Australia
\newline
}

\email{omar.foda@unimelb.edu.au}

\address{
\!\!\!\!\!\!\!\!\!
$^{{\scriptstyle 2}}$
Max-Planck-Institut f\"ur Mathematik, Vivatsgasse 7,
53111 Bonn, Germany
\newline
}

\email{masahidemanabe@gmail.com}

	
\begin{abstract}
We show, in a number of simple examples, that Macdonald-type $qt$-deformations 
of topological string partition functions are equivalent to topological string 
partition functions that are without $qt$-deformations but with brane condensates, 
and that these brane condensates lead to geometric transitions. 
\end{abstract}
	
\maketitle

%\tableofcontents

\section{Introduction}
\label{section.01}
%%%%%%%%%%%%%%%%%%%%%%%%%%%%%%%%%%%%%%%%%%%%%%%%%%%%%%%%%%%%%%%
%%%%%%%%%%%%%%%%%%%%%%%%%%%%%%%%%%%%%%%%%%%%%%%%%%%%%%%%%%%%%%%
\noindent \textit{We recall the topological vertex, its refinements and deformations, 
and ask what the physical interpretation of a specific Macdonald-type deformation is.}
\medskip 

\subsection{A hierarchy of topological vertices}
\label{subsec:hierarchy}

\subsubsection{Abbreviations} To simplify the presentation, we use  
\1 
\textit{string, string partition function, vertex, etc.} for 
\textit{topological string, topological string partition function, topological vertex, etc.},
which should cause no confusion, as we only consider the latter, 
and use \textit{topological} only for emphasis when that is needed, 
\2  
\textit{$qt$-string partition function, $qt$-quantum curve, etc.} for 
\textit{$qt$-deformed string partition function, $qt$-deformed quantum curve, etc.} 
\3  
\textit{refined} as in \textit{refined partition functions, etc.}, when discussing objects 
that are refined in the sense of 
\cite{awata.kanno.01, awata.kanno.02, iqbal.kozcaz.vafa};
otherwise, no refinement should be inferred, 
and \textit{unrefined} is used only for emphasis when that is needed,
\4  
\textit{the $qt$-version of $\cdots$} for the version of an object that is deformed in the sense
of \cite{vuletic, foda.wu.02}, and
\5 
\textit{a brane condensate}, or simply \textit{a condensate} is a set of infinitely-many 
brane insertions. 

\subsubsection{The original vertex as a normalized 1-parameter generating function of plane 
partitions with fixed asymptotic boundaries} 
In \cite{iqbal}, Iqbal introduced a systematic way to compute $A$-model string partition 
functions in terms of gluing copies of a trivalent topological vertex, and constructed 
a special case of that vertex where one of the three legs is trivial. 
In \cite{aganagic.klemm.marino.vafa}, Aganagic, Klemm, Mari\~no and Vafa constructed the 
full topological vertex $\cC_{\, Y_{\, 1} \, Y_{\, 2} \, Y_{\, 3}} \ll x \rr$, where all 
legs are non-trivial, that we refer to in the present work as \textit{the original vertex}\footnote{\,
To streamline the presentation, we make a number of departures from conventional notation. 
We state these changes as we introduce them, and list them in section \ref{notational.changes}. 
In particular, we use $x$, instead of $q$, for the weight of a box in 
$\cC_{\, Y_{\, 1} \, Y_{\, 2} \, Y_{\, 3}} \ll x \rr$.
}. 
It depends on a single parameter $x$, and a set of three Young diagrams, $Y_{\, 1}$, $Y_{\, 2}$ 
and $Y_{\, 3}$, and has a combinatorial interpretation as a normalized partition function of 3D 
plane partitions \cite{okounkov.reshetikhin.vafa}, where each box in each plane partition is 
assigned a weight $x$. All plane partitions generated by 
$\cC_{\, Y_{\, 1} \, Y_{\, 2} \, Y_{\, 3}} \ll x \rr$ satisfy fixed asymptotic boundary conditions 
specified by $Y_{\, 1}$, $Y_{\, 2}$ and $Y_{\, 3}$. Copies of 
$\cC_{\, Y_{\, 1}\, Y_{\, 2}\, Y_{\, 3}} \ll x \rr$ can be glued to form string partition functions. 
Using geometric engineering \cite{katz.klemm.vafa, katz.mayr.vafa}, these string partition functions 
are identified with instanton partition functions in 5D supersymmetric gauge theories on 
$\RR^4 \times S^1$, in a self-dual $\Omega$-background with Nekrasov parameters 
$\eo \, + \, \et \, = \, 0$ \cite{nekrasov,nekrasov.okounkov}.
Using the AGT/W correspondence \cite{alday.gaiotto.tachikawa, wyllard}, the 4D limit of these 5D 
instanton partition functions are identified with conformal blocks in 2D conformal field theories 
with an integral central charge $c$. 

\subsubsection{The refined vertex as a normalized 2-parameter generating function of plane 
partitions with fixed asymptotic boundaries}
In \cite{awata.kanno.01, awata.kanno.02}, Awata and Kanno introduced a refined version of 
$\cC_{\, Y_{\, 1} \, Y_{\, 2} \, Y_{\, 3}} \ll x \rr$, and in \cite{iqbal.kozcaz.vafa}, Iqbal, 
Kozcaz and Vafa introduced yet another refined version of the same object. 
In \cite{awata.feigin.shiraishi}, Awata, Feigin and Shiraishi proved that these two refinements 
are equivalent. In the present work, we focus on \textit{the refined vertex} 
$\cR_{\, Y_{\, 1}\, Y_{\, 2}\, Y_{\, 3}} \ll x, y\rr$ of \cite{iqbal.kozcaz.vafa}.\footnote{\,
We use $\ll x, y \rr$ instead of $\ll q, t\rr$ for the parameters, and 
$\cR_{\, Y_{\, 1}\, Y_{\, 2}\, Y_{\, 3}} \ll x, y\rr$ instead of 
$\cC_{\, Y_{\, 1}\, Y_{\, 2}\, Y_{\, 3}} \ll t, q\rr$ for the refined vertex of 
\cite{iqbal.kozcaz.vafa}. We reserve the parameters $\ll q, t\rr$ for the Macdonald-type 
deformation parameters of \cite{vuletic, foda.wu.02} introduced in section 
\textbf{\ref{mac.top.ver.introduced}}.}
It depends on two parameters $\ll x, y\rr$, and a set of three Young diagrams, 
$Y_{\, 1}$, $Y_{\, 2}$ and $Y_{\, 3}$, and has a combinatorial interpretation as a normalized 
partition function of 3D plane partitions. Each box in each plane partition is assigned a weight 
$x$ or $y$ as follows. One splits each plane partition diagonally into vertical Young diagrams. 
Scanning the vertical Young diagrams from one end to the other, a box in a plane partition is 
assigned a weight $x$ if it belongs to a vertical Young diagram that protrude with respect to 
the preceding Young diagram, and a weight $y$ if it belongs to a vertical Young diagram that 
does not. All plane partitions generated by $\cR_{\, Y_{\, 1}\, Y_{\, 2}\, Y_{\, 3}} \ll x, y\rr$ 
satisfy fixed asymptotic boundary conditions specified by $Y_{\, 1}$, $Y_{\, 2}$ and $Y_{\, 3}$. 
Copies of $\cR_{\, Y_{\, 1}\, Y_{\, 2}\, Y_{\, 3}} \ll x, y \rr$ can be glued to form refined  
string partition functions. Using geometric engineering \cite{katz.klemm.vafa, katz.mayr.vafa}, 
these refined string partition functions are identified with instanton partition functions in 
5D supersymmetric gauge theories on $\RR^4 \times S^1$, in a generic $\Omega$-background, with 
Nekrasov parameters $\eo \, + \, \et \, \neq \, 0$ \cite{nekrasov, nekrasov.okounkov}. Using 
the AGT/W correspondence \cite{alday.gaiotto.tachikawa, wyllard}, the 4D limits of these 5D 
instanton partition functions are identified with conformal blocks in 2D conformal field 
theories with a non-integral central charge $c$.

\subsubsection{The Macdonald vertex as a $qt$-deformation of the refined vertex} 
\label{mac.top.ver.introduced}
In \cite{vuletic}, Vuleti\'c introduced a deformation of MacMahon's generating function of 
plane partitions, in terms of two Macdonald-type parameters $\ll q, t\rr$. 
This deformation is independent 
of the refinement introduced in \cite{awata.kanno.01, awata.kanno.02} and \cite{iqbal.kozcaz.vafa}, 
as one can check by considering 
$\cR^{\, \prime}_{\, \emptyset \, \emptyset \, \emptyset} \ll x, y \rr$, the unnormalized version of
$\cR^{         }_{\, \emptyset \, \emptyset \, \emptyset} \ll x, y \rr$, which is a refinement of 
MacMahon's generating function, but is different from that of \cite{vuletic}. 
In \cite{foda.wu.02}, $\cR_{\, Y_{\, 1}\, Y_{\, 2}\, Y_{\, 3}} \ll x, y \rr$ was deformed using 
the same Macdonald-type parameters $\ll q, t\rr$ that were used in \cite{vuletic}, to obtain 
\textit{the Macdonald vertex} $\cM_{\, Y_{\, 1} \, Y_{\, 2} \, Y_{\, 3}}^{\, q \, t} \ll x, y \rr$.\footnote{\,
We call the ratio $x/y$ \textit{a refinement}, and in the limit $x \rightarrow y$, 
the refined vertex reduces to the original one, and 
we call the ratio $q/t$ \textit{a deformation}, and in the limit $q \rightarrow t$, 
the Macdonald vertex reduces 
to the original vertex, for $x=y$, or to the refined vertex, for $x \neq y$.}
Copies of $\cM_{\, Y_{\, 1} \, Y_{\, 2} \, Y_{\, 3}}^{\,q\,t} \ll x, y \rr$ can be glued to form 
$qt$-string partition functions that are 5D $qt$-instanton partition functions. The latter have 
well-defined 4D-limits and, for generic values of $\ll q, t\rr$, contain infinite towers of poles 
for every pole that is present in the limit $q \rightarrow t$ \cite{foda.wu.02}.

\subsubsection{Limits of the Macdonald vertex} In constructing the original and the refined vertex, 
(undeformed) free bosons that satisfy the Heisenberg algebra,

\begin{equation} 
\left[a_m, a_n\right] = n \, \delta_{m + n, \, 0} \, , 
\end{equation}

\noindent play a central role \cite{okounkov.reshetikhin.vafa, iqbal.kozcaz.vafa}. Similarly, 
in constructing the Macdonald vertex, $qt$-free bosons that satisfy the $qt$-Heisenberg algebra,

\begin{equation} 
\left[a_m^{\,q \,t}, a_n^{\,q \,t}\right] = 
n \, \ll \frac{1 - q^{\, |n|}}{1 - t^{\, |n|}}\rr \delta_{m + n, \, 0} \, , 
\end{equation}

\noindent play a central role. In the limit $q \rightarrow t$, 
$\cM_{\, Y_{\, 1} \, Y_{\, 2} \, Y_{\, 3}}^{\,q\,t} \ll x, y \rr 
\rightarrow \,
\cR_{\, Y_{\, 1}\, Y_{\, 2}\, Y_{\, 3}} \ll x, y \rr$,  
and in the limit $x \rightarrow y$, 
$\cM_{\, Y_{\, 1} \, Y_{\, 2} \, Y_{\, 3}}^{\,q\,t} \ll x, y \rr 
\rightarrow\, 
\cC_{\, Y_{\, 1} \, Y_{\, 2} \, Y_{\, 3}}^{\,q\,t} \ll x \rr$, 
which is a $qt$-deformation of $\cC_{\, Y_{\, 1} \, Y_{\, 2} \, Y_{\, 3}} \ll x \rr$.

\subsection{The physical interpretation of the $qt$-deformation}
It is clear by inspection of explicit computations that the Macdonald parameter ratio $q/t$ is 
a different object from either the $M$-theory circle radius $R$ or the refinement parameter ratio $x/y$.\footnote{\,
One can also introduce an elliptic nome $p$ \cite{iqbal.hollowood.vafa, iqbal.kozcaz.yau, zhu, foda.zhu}, 
which is yet another parameter. In section \ref{4.parameters}, we discuss what we know about the interpretation 
of the four parameters, $R$, $x/y$, $q/t$, and $p$.} The purpose of the present work is to shed light on the 
geometric and/or physical interpretation of the $qt$-deformation. To do this, we consider simple string 
partition functions, and show that in $M$-theory terms, the deformation $q/t \neq 1$ describes a condensation 
of $M5$-branes that lead to geometric transitions that change the topology of the original Calabi-Yau 3-fold
\cite{gopakumar.vafa.01}. In conformal field theory terms, we expect that it describes a condensation of vertex
operators that push the conformal field theory off criticality \cite{zamolodchikov}.

\subsection{Outline of contents}
In section \textbf{\ref{section.02}},   
we include comments on notation used in the text, and definitions of combinatorial objects, 
including MacMahon's generating function of plane partitions, its refinement and $qt$-deformation, and
in \textbf{\ref{section.03}},  
include basic facts related to the original topological vertex, the refined topological vertex, 
and their $qt$-deformations.  
In section \textbf{\ref{section.04}},  
we give our first example of the equivalence of $qt$-deformation and brane condensation, 
which shows that 
the refined $qt$-string partition function on $\CC^{\, 3}$ is equivalent to 
a refined string partition function on $\CC^{\, 3}$ 
with no $qt$-deformation but in the presence of condensates, 
and in \textbf{\ref{section.05}}, we give our second example, which shows that 
a refined $qt$-deformed partition function on $\CC^{\, 3}$ with a single-brane insertion is equivalent to 
its counterpart (also with a single-brane insertion) 
with no $qt$-deformation but in the presence of condensates.
In section \textbf{\ref{section.06}},  
we discuss the relation of the condensates and geometric transitions 
in the context of unrefined objects, and 
in \textbf{\ref{section.07}}, 
we discuss the $qt$-quantum curves associated with $qt$-partition function.
Finally, in section \textbf{\ref{section.8}},  
we collect a number of remarks, and discuss the various parameters that can 
appear in topological vertices and the relation with conformal field theory, and 
in appendix \textbf{\ref{app_formula}}, we collect useful skew Schur function 
identities that are used freely in the text. 

%%%%%%%%%%%%%%%%%%%%%%%%%%%%%%%%%%%%%%%%%%%%%%%%%%%%%%%%%%%%%%%
%%%%%%%%%%%%%%%%%%%%%%%%%%%%%%%%%%%%%%%%%%%%%%%%%%%%%%%%%%%%%%%

\section{Notation and definitions}
\label{section.02}
%%%%%%%%%%%%%%%%%%%%%%%%%%%%%%%%%%%%%%%%%%%%%%%%%%%%%%%%%%%%%%%
%%%%%%%%%%%%%%%%%%%%%%%%%%%%%%%%%%%%%%%%%%%%%%%%%%%%%%%%%%%%%%%
\noindent \textit{We collect comments on notation, definitions of combinatorial objects, including 
variations on MacMahon's generating function of plane partitions that appear in the sequel.}
\medskip

\subsection{Notation}

\subsubsection{Deviations from standard notation}
\label{notational.changes}
We use the variables $\ll x, y\rr$ as box weights/refinement parameters, instead of the variables 
$\ll q, t\rr$ used in \cite{awata.kanno.01, awata.kanno.02, iqbal.kozcaz.vafa}. We use
$\cR_{\, Y_{\, 1}\, Y_{\, 2}\, Y_{\, 3}} \ll x, y \rr$ for the refined vertex instead of 
$\cC_{\, Y_{\, 1}\, Y_{\, 2}\, Y_{\, 3}} \ll t, q \rr$ as used in \cite{iqbal.kozcaz.vafa}.\footnote{\, 
See section \textbf{\ref{more.on.notation}} for a more detailed relation.} 
We reserve the variables $\ll q, t\rr$ for the Macdonald-type deformation parameters that appear 
in the Macdonald  vertex $\cM_{\, Y_{\, 1} \, Y_{\, 2} \, Y_{\, 3}}^{\,q\,t} \ll x, y \rr$ of 
\cite{foda.wu.02}. 

\subsubsection{Sets}
\label{section.sets}
$\rho$ is the set of negative half-integers     $\ll \rho_{\, 1}, \rho_{\, 2}, \ldots \rr$ with $\rho_i = - i + 1/2$, 
that is $\ll \rho_{\, 1}, \rho_{\, 2}, \ldots \rr = \ll - \, 1/2, - \, 3/2, \ldots \rr$,  
and 
$\bi$  is the set of non-zero positive integers $\ll      1,      2, \ldots \rr$.

\subsection{Combinatorics}

\subsubsection{Cells in the lower-right quadrant}
Consider the lower-right quadrant in $\RR^{\, 2}$, bounded by the right-half of the $x$-axis and the lower-half 
of the $y$-axis. The intersection point of the $x$- and $y$-axes to be the origin with coordinates
$\ll 0, 0\rr$, the $x$-coordinate increases to the right, the $y$-coordinate increases downwards. 
We divide this quadrant into cells of unit-length in each direction.
A cell $\ws$ has coordinates $\ll i, j \rr$, if the coordinates of the lower-right corner of the cell are 
$\ll i, j \rr$. 

\subsubsection{Young diagrams}
\label{section.young.diagrams}
$Y$ is a Young diagram in the lower-right quadrant of $\RR^{\, 2}$ that consists of rows of cells of positive 
integral lengths $y_{\, 1} \geqslant y_{\, 2} \geqslant \cdots \geqslant 0$, and 
$Y^{\, \prime}$ is the transpose of $Y$ that consists of rows of cells of positive integral lengths 
$y_{\, 1}^{\, \prime} \geqslant y_{\, 2}^{\, \prime} \geqslant \cdots \geqslant 0$.
$y_{\, 1}^{\, \prime}$ is the number of (non-zero) parts in $Y$.
The infinite profile of $Y$ consists of the union of 
\1 a semi-infinite line that extends from right to left along the positive, right-half of the $x$-axis, 
from $x = \infty$ to $x = y_{\, 1}$, 
\2 the finite profile of $Y$, and 
\3 a semi-infinite line that extends from top to bottom along the positive, lower-half of the $y$-axis, 
from $y = y_{\, 1}^{\, \prime}$ to $y = \infty$.\footnote{\,
In our notation, the positive half of $y$-axis is the lower-half that extends downwards.}

\subsubsection{Arms, legs and hook lengths}
Consider a Young diagram $Y$, and a cell $\wsq_{i j}$ with coordinates $\ll i, j \rr$ such that 
$\wsq_{i j}$ is \textit{not} necessarily inside $Y$. The           
arm $A^{}_{\, \wsq_{i j}}$, 
leg $L^{}_{\, \wsq_{i j}}$, 
extended arm $A^{+}_{\, \wsq_{i j}}$, 
extended leg $L^{+}_{\, \wsq_{i j}}$, and 
hook $H^{}_{\, \wsq_{i j}}$ of $\wsq_{i j}$, with respect to the infinitely-extended profile 
of $Y$, are, 

\begin{equation}
A_{\, \wsq_{i j}}  =  y_i             -\, j, 
\quad
L_{\, \wsq_{i j}}  =  y^{\, \prime}_j -\, i,
\quad 
A^+_{\, \wsq_{i j}}  =  A_{\, \wsq_{i j}} + 1, 
\quad
L^+_{\, \wsq_{i j}}  =  L_{\, \wsq_{i j}} + 1, 
\quad
H_{\, \wsq_{i j}}  = A_{\, \wsq_{i j}} + L_{\, \wsq_{i j}} + 1, 
\end{equation}

\noindent where $y^{\, \prime}_j$ is the length of the $j$-row in $Y^{\, \prime}$, 
which is the $j$-column in $Y$. We also define, 

\begin{equation}
\left| Y \right|=\sum_{\wsq \in Y} \, 1, 
\quad
\frac12 \left\Vert Y \right\Vert^{\, 2} = \sum_{\wsq \in Y} \ll  A_{\, \wsq}+\frac12 \rr, 
\quad
\frac12 \, \kappa_{\, Y}  = 
\frac12 
\ll 
\left\Vert Y             \right\Vert^{\, 2} - 
\left\Vert Y^{\, \prime} \right\Vert^{\, 2}
\rr 
=\sum_{\ll i, j \rr \in Y} \ll j - i \rr
\end{equation}

\subsection{The framing factor}
We use the notation $f_{\, Y} \ll x \rr$ for the framing factor of the original vertex
\cite{marino.vafa,aganagic.klemm.marino.vafa},

\begin{equation}
f_{\, Y} \ll x \rr = \ll -1\rr^{\left|Y \right|} \, x^{\frac12 \, \kappa_{\, Y}},
\label{framing_tv}
\end{equation}

\noindent and,

\begin{equation}
f_{\, Y} \ll x,y \rr = \ll -1 \rr^{\left|Y \right|} \,
x^{\, - \,   \frac12 \, \left\Vert Y^{\, \prime} \right\Vert^{\, 2}} \,
y^{\, \frac12 \, \left\Vert Y             \right\Vert^{\, 2}},
\label{framing_ref_tv}
\end{equation}

\noindent for the refined framing factor introduced in \cite{taki.01} of the refined vertex. 

\subsection{Splitting indices}
\label{splitting.indices}
Starting from a sequence $\textbf{a} = \ll a_1, a_2, \ldots \rr$, one can split the single index $I$ 
of any element $a_I$ into two indices $ij$, so that $a_I \to a_{ij}$. One way to split the indices is
in the following example. 

\subsubsection{Example} 
We proceed in two steps. 
\1 Position the elements of the 1-dimensional sequence $\ll a_1, a_2, \ldots \rr$ along the anti-diagonals
of a 2-dimensional array, as in, 

\begin{equation} 
\ll a_1, a_2, \ldots \rr 
\quad 
\mapsto
\quad 
\begin{array}{cccc} 
a_1    & a_2    & a_4 & \cdots \\ 
a_3    & a_5    &     &        \\
a_6    &        &     &        \\
\end{array} 
\label{splitting.indices.equation.01}
\end{equation}

\noindent
\2 Map the array with single-index elements to an array with double-index elements, where the double-indices 
are in conventional order, as in,  

\begin{equation} 
\begin{array}{cccc} 
a_1 & a_2 & a_4     & \cdots \\ 
a_3 & a_5 &         &        \\ 
a_6 &     &         &        \\
\end{array}
\quad
\mapsto 
\quad 
\begin{array}{cccc} 
a_{11} & a_{12} & a_{13}  & \cdots \\ 
a_{21} & a_{22} &         &        \\ 
a_{31} &        &         &        \\
\end{array}
\label{splitting.indices.equation.02}
\end{equation}

\noindent Any such splitting of indices is far from unique. However, if the splitting rule is well-defined, 
as in \eqref{splitting.indices.equation.01}--\eqref{splitting.indices.equation.02}, then it is bijective,  
and all such splittings are in bijection \textit{via} the original 1-dimensional sequence. 

\subsection{Variations on MacMahon's generating functions}

\subsubsection{Notation} To streamline the notation, we use the redundant notation 
$M_{\, x \, x}$ for MacMahon's original generating function of plane partitions, so that we can write
$M_{\, x \, y}$ for its refined counterpart, and 
$M_{\, x \, y}^{\, q \, t}$ for the $qt$-version of the latter.
$M_{\, x \, x}^{\, q \, t}$ is the $qt$-MacMahon generating function of Vuleti\'c, and
$M_{\, x \, y}^{\, q \, q} = M_{\, x \, y}$.

\subsubsection{$M_{\, x \, x}$} 
MacMahon's generating function of plane partitions is, 

\begin{equation}
M_{\, x \, x} = \prod_{m \, = \, 1}^{\infty} \ll \frac{1}{1-x^{\, m}} \rr^{\, m}
= \exp{\ll \sum_{n\, =\, 1}^{\infty}\frac{1}{n \ll x^{\, n/2} - x^{\, - \, n/2} \rr^{\, 2}}\rr}
\label{macmahon}
\end{equation}

\noindent The first equation in (\ref{macmahon}) is the definition of the MacMahon generating function. 
The second is obtained by direct expansion of the logarithms of both sides. 
All refined and $qt$-versions of this equation, in the sequel, are proven similarly.

\subsubsection{$M_{\, x \, y}$} 
The refined MacMahon's generating function of plane partitions is \cite{iqbal.kozcaz.vafa}, 

\begin{equation}
M_{\, x \, y} = 
\prod_{m, n = 1}^{\infty} \frac{1}{1 - x^{\, m} y^{\, n - 1}} =
\exp{\ll \sum_{n\, =\, 1}^{\infty}\frac{\ll x/y \rr^{\, n/2}}{n \ll x^{\, n/2} - x^{\, - \, n/2} \rr
\ll y^{\, n/2} - y^{\, - \, n/2} \rr}\rr}
\label{ref_macmahon}
\end{equation}

\noindent In the limit $y \rightarrow x$, $M_{\, x \, y} \rightarrow M_{\, x \, x}$. 

\subsubsection{$M_{\, x \, x}^{\, q \, t}$}
The $qt$-version MacMahon's generating function of plane partitions is,

\begin{equation}
M_{\, x \, x}^{\, q \, t} =
\prod_{i \, = \, 0}^{\infty} \prod_{m \, = \, 1}^{\infty} 
\ll \frac{1-q^{\, i} t x^{\, m}}{1- q^{\, i} x^{\, m}} \rr^{\, m} =
\exp{
\ll 
\sum_{n\, =\, 1}^{\infty}
\frac{1}{n \ll x^{\, n/2} - x^{\, - \, n/2} \rr^{\, 2}}
\ll \frac{1 - t^n}{1 - q^n} \rr 
\rr
}
\label{vuletic_mac}
\end{equation}

\noindent which is the $qt$-MacMahon generating function introduced by Vuleti\'c in \cite{vuletic}.
In the limit 
$t \rightarrow q$, $M_{\, x \, x}^{\, q \,t} \rightarrow M_{\, x \, x}^{\, q \,q} = M_{\, x \, x}$. 

\subsubsection{$M_{\, x \, y}^{\,q \,t}$}
The refined $qt$-version MacMahon's generating function of plane partitions 
is \cite{foda.wu.02},

\begin{equation}
M_{\, x \, y}^{\,q \,t} = 
\prod_{i \, = \, 0}^{\infty} \prod_{m, n = 1}^{\infty} 
\frac{1 - q^{\, i} \, t \, x^{\, m} \, y^{\, n - 1}}{1 - q^{\, i} x^{\, m} y^{\, n - 1}} = 
\exp{ 
\ll 
\sum_{n \, =  \, 1}^{\infty}
\frac{
\ll x / y \rr^{\, n/2}
}{
n \ll x^{\, n/2} - x^{\, - \, n/2} \rr \ll y^{\, n/2} - y^{\, - \, n/2} \rr
}
\ll \frac{1 - t^n}{1 - q^n} \rr 
\rr
}
\label{def_ref_macmahon}
\end{equation}

\noindent In the limit $y \rightarrow x$, $M_{\, x \, y}^{\, q \,t} = M_{\, x \, x}^{\, q \, t}$, and so on.

%%%%%%%%%%%%%%%%%%%%%%%%%%%%%%%%%%%%%%%%%%%%%%%%%%%%%%%%%%%%%%%
%%%%%%%%%%%%%%%%%%%%%%%%%%%%%%%%%%%%%%%%%%%%%%%%%%%%%%%%%%%%%%%

\section{Topological vertices}
\label{section.03}
%%%%%%%%%%%%%%%%%%%%%%%%%%%%%%%%%%%%%%%%%%%%%%%%%%%%%%%%%%%%%%%
%%%%%%%%%%%%%%%%%%%%%%%%%%%%%%%%%%%%%%%%%%%%%%%%%%%%%%%%%%%%%%%
\noindent \textit{We recall basic facts related to the topological vertices introduced in 
section \textbf{\ref{subsec:hierarchy}}.}
\medskip 
\subsection{The original vertex of \cite{aganagic.klemm.marino.vafa}} 
\begin{figure}[t]
\centering
\includegraphics[width=105mm]{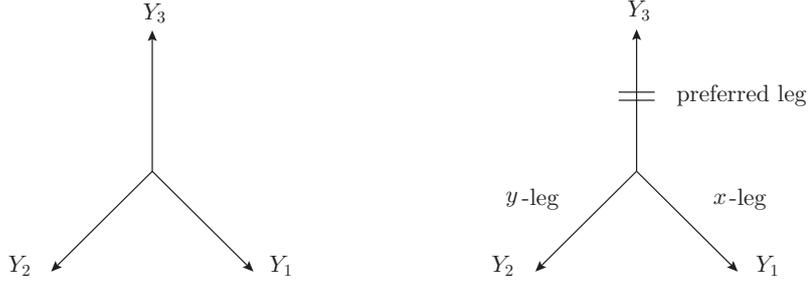}
\caption{The figure on the left represents the  vertex $\cC_{\, Y_{\, 1}\, Y_{\, 2}\, Y_{\, 3}} \ll x \rr$, and 
this vertex has the cyclic symmetry 
$\cC_{\, Y_{\, 1}\, Y_{\, 2}\, Y_{\, 3}} \ll x \rr =
 \cC_{\, Y_{\, 3}\, Y_{\, 1}\, Y_{\, 2}} \ll x \rr =
 \cC_{\, Y_{\, 2}\, Y_{\, 3}\, Y_{\, 1}} \ll x \rr$. 
The figure on the right represents the refined vertex 
$\cR_{\, Y_{\, 1} \, Y_{\, 2}\, Y_{\, 3}} \ll x, y \rr$ and the Macdonald vertex 
$\cM_{\, Y_{\, 1} \, Y_{\, 2} \, Y_{\, 3}}^{\,q \,t} \ll x, y \rr$. 
These two vertices break the cyclic symmetry and have the preferred leg. 
Note that $\cR_{\, Y_{\, 1} \, Y_{\, 2} \, Y_{\, 3}} \ll x, x \rr = \cC_{\, Y_{\, 1} \, Y_{\, 2} \, Y_{\, 3}^{\, \prime}} \ll x \rr$.}
\label{top_v}
\end{figure}

With reference to the 
figure on the left in \figref{top_v},
the normalized version of the original vertex
\footnote{\,
In the present work, we use $x$ for the weight of a box in a plane partition, instead of $q$ 
in \cite{iqbal, aganagic.klemm.marino.vafa}. For a review of the original vertex, see \cite{marino.review}.} 
of \cite{aganagic.klemm.marino.vafa} is,

\begin{multline}
\cC_{\, Y_{\, 1} \, Y_{\, 2} \, Y_{\, 3}} \ll x \rr =
x^{\, \frac12 \, \kappa_{Y_{\, 1}}} \, s_{\, Y_{\, 3}} \ll \, x^{\, \rho} \rr 
\sum_{\, Y} 
s_{\,Y_{\, 1}^{\, \prime} / Y} \ll x^{\, \rho \, + \, Y_{\, 3}            } \rr 
s_{\,Y_{\, 2}             / Y} \ll x^{\, \rho \, + \, Y_{\, 3}^{\, \prime}} \rr
\\
=
\ll -1 \rr^{\, \left| Y_{\, 2} \right| \, + \, \left| Y_{\, 3} \right|} \, f_{\, Y_{\, 1}} \ll x \rr
x^{\, \frac12 \, \left\Vert Y_{\, 3} \right\Vert^{\, 2}}
\ll \prod_{\wsq \in Y_{\, 3}} \frac{1}{1 - x^{\, H_{\, \wsq}}} \rr
\sum_{\, Y} 
s_{\, Y_{\, 1} / Y} \ll x^{\, - \, \rho \, - \, Y_{\, 3}} \rr s_{\, Y_{\, 2}^{\, \prime}/Y}\ll x^{\,- \, \rho \, - \, Y_{\, 3}^{\, \prime}} \rr
\label{aganagic.klemm.marino.vafa_vertex}
\end{multline} 

\noindent Here $x^{\,\rho \, + \, Y} = \ll x^{\,\rho_{\, 1} \, + \, y_{\, 1}}, x^{\,\rho_{\, 2} \, + \, y_{\, 2}}, \ldots \rr$, 
$x=\mathrm{e}^{\, - \, g_s}$, where $g_s$ is the string coupling constant, and $s_{\, Y_{\, 1} / Y_{\, 2}} \ll \mathbf{x} \rr$ 
is the skew Schur function defined in terms of a pair of Young diagrams $\ll Y_{\, 1}, Y_{\, 2} \rr$ and a set of possibly 
infinitely-many variables $\mathbf{x} = \ll x_1, x_2, \ldots \rr$. In the second equality, we have used the notation 
$f_{\, Y} \ll x \rr$ for the framing factor (\ref{framing_tv}), and the identities in appendix \textbf{A}.  

\subsubsection{Normalization} 
$\cC_{\, Y_{\, 1} \, Y_{\, 2} \, Y_{\, 3}} \ll x \rr$ is normalized by $M_{\, x \, x}$ such that 
$\cC_{\, \emptyset \, \emptyset \, \emptyset} \ll x \rr=1$. The unnormalized version is, 

\begin{equation}
\cC^{\, \prime}_{\, Y_{\, 1} \, Y_{\, 2} \, Y_{\, 3}} \ll x \rr = 
M_{\, x \, x} \, \cC_{\, Y_{\, 1} \, Y_{\, 2} \, Y_{\, 3}} \ll x \rr
\end{equation}

\subsubsection{$\cC^{\, \prime}_{\, Y_{\, 1}\, Y_{\, 2}\, Y_{\, 3}} \ll x \rr$ and $M_{\, x \, x}$ 
as partition functions}
The unnormalized vertex $\cC^{\, \prime}_{\, Y_{\, 1}\, Y_{\, 2}\, Y_{\, 3}} \ll x \rr$ 
is the open topological $A$-model partition function on $\CC^{\, 3}$ with three special 
Lagrangian submanifolds. 
$M_{\, x \, x}$ is the closed topological $A$-model partition function on $\CC^{\, 3}$. 
The figure on the left in \figref{top_v} is the toric web diagram of $\CC^{\, 3}$.

\subsubsection{Choice of framing} One can choose the framing of 
$\cC_{\, Y_{\, 1}\, Y_{\, 2}\, Y_{\, 3}} \ll x \rr$ as, 

\begin{equation}
\cC_{\, Y_{\, 1}\, Y_{\, 2}\, Y_{\, 3}} \ll x \rr \quad \to \quad 
\ll \prod_{\, i \, = \, 1, \, 2, \, 3} f_{\, Y_i}\ll x \rr^{f_i} \rr
\cC_{\, Y_{\, 1}\, Y_{\, 2}\, Y_{\, 3}} \ll x \rr, \quad f_1, f_2, f_3 \in \ZZ, 
\label{aganagic.klemm.marino.vafa_frame}
\end{equation}

\noindent where $f_{\, Y} \ll x \rr$ is the framing factor (\ref{framing_tv}).

\subsection{The refined vertex of \cite{iqbal.kozcaz.vafa}}
With reference to the figure on the right in \figref{top_v}, the normalized refined vertex of \cite{iqbal.kozcaz.vafa} is,

\begin{multline}
\cR_{\, Y_{\, 1}\, Y_{\, 2}\, Y_{\, 3}} \ll x, y\rr =
\ll -1 \rr^{\, \left|Y_{\, 2} \right| \, + \, \left|Y_{\, 3} \right|} \, f_{\, Y_{\, 1}} \ll x, y \rr
x^{\frac12 \left\Vert Y_{\, 3}^{\, \prime} \right\Vert^{\, 2}}
\ll \prod_{\wsq \in Y_{\, 3}} \frac{1}{1 - x^{\, L^+_{\, \wsq}} y^{\, A_{\, \wsq}}} \rr \, \times
\\ 
\sum_Y  
\ll \frac{y}{x} \rr^{\frac12 \ll | Y | - | Y_{\, 1} | + | Y_{\, 2} | \rr}
s_{\, Y_{\, 1}^{         } / Y} \ll y^{\, - \rho} x^{\, - Y_{\, 3}^{\, \prime}} \rr
s_{\, Y_{\, 2}^{\, \prime} / Y} \ll x^{\, - \rho} y^{\, - Y_{\, 3}            } \rr,
\label{iqbal.kozcaz.vafa_vertex}
\end{multline}

\noindent where $f_{\, Y} \ll x,y \rr$ is the refined framing factor 
(\ref{framing_ref_tv}). In the limit $y \rightarrow x$,

\begin{equation}
\cR_{\, Y_{\, 1}\, Y_{\, 2}\, Y_{\, 3}} \ll x, y \rr 
\quad \rightarrow \quad
\cC_{\, Y_{\, 1}\, Y_{\, 2}\, Y_{\, 3}^{\, \prime}} \ll x    \rr
\label{rel_CR}
\end{equation}

\subsubsection{Remark.} The dependence on the Young diagram $Y_{\, 3}$ in 
$\cR_{\, Y_{\, 1}\, Y_{\, 2}\, Y_{\, 3}} \ll x, y \rr$ on the left hand side of \eqref{rel_CR} 
is replaced by a dependence on its transpose $Y_{\, 3}^{\, \prime}$ in  
$\cC_{\, Y_{\, 1}\, Y_{\, 2}\, Y_{\, 3}^{\, \prime}} \ll x \rr$ on the right hand side. 

\subsubsection{Remark.}
\label{more.on.notation}
$\ll t, q\rr$ in \cite{iqbal.kozcaz.vafa} become $\ll x, y\rr$ in the present work, and the refined vertex 
$\cC_{\, Y_{\, 1}\, Y_{\, 2}\, Y_{\, 3}} \ll t, q \rr$ in \cite{iqbal.kozcaz.vafa} is related to 
$\cR_{\, Y_{\, 1}\, Y_{\, 2}\, Y_{\, 3}} \ll x, y\rr$ in the present work by,

\begin{equation}
\cC_{\, Y_{\, 1}\, Y_{\, 2}\, Y_{\, 3}} \ll t, q \rr =  
\ll -1 \rr^{|Y_{\, 1}| + |Y_{\, 2}|} \, f_{\, Y_{\, 3}} \ll x, y \rr
\cR_{\, Y_{\, 2}\, Y_{\, 1}\, Y_{\, 3}} \ll x, y\rr 
\label{fn:ref_v}
\end{equation}

\subsubsection{Choice of framing} One can choose the framing of 
$\cR_{\, Y_{\, 1}\, Y_{\, 2}\, Y_{\, 3}} \ll x, y \rr$ as, 

\begin{equation}
\cR_{\, Y_{\, 1}\, Y_{\, 2}\, Y_{\, 3}} \ll x, y \rr \quad \to \quad 
\ll \prod_{i=1,2,3} f_{\, Y_i} \ll x,y \rr^{f_i} \rr
\cR_{\, Y_{\, 1}\, Y_{\, 2}\, Y_{\, 3}} \ll x, y \rr, \quad f_1, f_2, f_3 \in \ZZ
\label{iqbal.kozcaz.vafa_frame}
\end{equation}

\subsubsection{Normalization} 
$\cR_{\, Y_{\, 1} \, Y_{\, 2} \, Y_{\, 3}} \ll x, y \rr$ is normalized by $M_{\, x \, y}$ such that 
$\cR_{\, \emptyset \, \emptyset \, \emptyset} \ll x \rr=1$. The unnormalized version is, 

\begin{equation}
\cR^{\, \prime}_{\, Y_{\, 1} \, Y_{\, 2} \, Y_{\, 3}} \ll x, y \rr = 
M_{\, x \, y} \, \cR_{\, Y_{\, 1} \, Y_{\, 2} \, Y_{\, 3}} \ll x, y \rr
\label{norm_ref_v}
\end{equation}

\subsection{The Macdonald vertex of \cite{foda.wu.02}}
With reference to the figure on the right in \figref{top_v}, the normalized Macdonald vertex 
of \cite{foda.wu.02} is, 

\begin{equation}
\cM^{ \, q \, t}_{\, Y_{\, 1} \, Y_{\, 2} \, Y_{\, 3}} \ll x, y \rr =
\ll 
\prod_{i \, = \, 0}^{\infty}\prod_{\wsq \in Y_{\, 3}}
\frac{
1 - q^{\, i} \, t \, x^{\, L^+_{\, \wsq}} y^{\, A_{\wsq}}
}{
1 - q^{\, i} \,      x^{\, L^+_{\, \wsq}} y^{\, A_{\wsq}}} 
\rr
\sum_Y  
P_{\, Y_{\, 1} / Y}^{\, q \, t} \ll y^{\, \bi - 1} x^{\, - \, Y_{\, 3}^{\, \prime}} \rr 
Q_{\, Y_{\, 2} / Y}^{\, q \, t} \ll x^{\, \bi} y^{\, -Y_{\, 3}} \rr
\label{qt_def_vertex}
\end{equation}

\noindent Here 
$P_{\, Y_{\, 1} / Y_{\, 2}}^{\, q \, t}\ll \mathbf{x}\rr$ and 
$Q_{\, Y_{\, 1} / Y_{\, 2}}^{\, q \, t}\ll \mathbf{x}\rr$ 
are the skew Macdonald and dual Macdonald functions defined 
for a pair of Young diagrams $\ll Y_{\, 1}, Y_{\, 2} \rr$ and a set 
of possibly infinitely-many variables $\mathbf{x} = \ll x_1, x_2, \ldots \rr$. 

\subsubsection{Choice of framing} No choice of framing of 
$\cM^{\, q \, t}_{\, Y_{\, 1}\, Y_{\, 2}\, Y_{\, 3}} \ll x, y \rr$ was discussed in \cite{foda.wu.02}, 
and none will be needed in the present work.

\subsubsection{Normalization} 
$\cM^{ \, q \, t}_{\, Y_{\, 1} \, Y_{\, 2} \, Y_{\, 3}} \ll x, y \rr$ is normalized by $M^{\, q \, t}_{\, x \, x}$
such that $\cM^{ \, q \, t}_{\, \emptyset \, \emptyset \, \emptyset} \ll x, y \rr=1$. 
The unnormalized version is, 

\begin{equation}
\cM^{\, \prime \, q \, t}_{\, Y_{\, 1} \, Y_{\, 2} \, Y_{\, 3}} \ll x, y \rr = 
M^{\, q \, t}_{\, x \, y} \, 
\cM^{\, q \, t}_{\, Y_{\, 1} \, Y_{\, 2} \, Y_{\, 3}} \ll x, y \rr
\end{equation}

%%%%%%%%%%%%%%%%%%%%%%%%%%%%%%%%%%%%%%%%%%%%%%%%%%%%%%%%%%%%%%%
%%%%%%%%%%%%%%%%%%%%%%%%%%%%%%%%%%%%%%%%%%%%%%%%%%%%%%%%%%%%%%%

\section{A $qt$-partition function from brane condensates}
\label{section.04}
%%%%%%%%%%%%%%%%%%%%%%%%%%%%%%%%%%%%%%%%%%%%%%%%%%%%%%%%%%%%%%%
%%%%%%%%%%%%%%%%%%%%%%%%%%%%%%%%%%%%%%%%%%%%%%%%%%%%%%%%%%%%%%%
\noindent \textit{We give an example of a refined $qt$-deformed partition function 
that is obtained from its undeformed counterpart via brane condensation.}
\medskip

\subsection{From $M5$-branes to surface operators}
Consider $M$-theory on,  

\begin{equation}
\RR^4 \times S^1 \times X,
\end{equation}

\noindent where $S^1$ is the $M$-theory circle, and $X$ is a local toric Calabi-Yau 3-fold such that 
the topological $A$-model on $X$ geometrically engineers a 5D $SU(N)$ supersymmetric gauge theory on 
$\RR^4 \times S^1$ with $\ll \CC^{\, \times}\rr^{\, 2}$-equivariant parameters $x, y$ acting on $\RR^4$ ($\Omega$-background) \cite{katz.klemm.vafa, katz.mayr.vafa}. We introduce $M5$-branes 
on the submanifold,   

\begin{equation}
\RR^{\, 2} \times S^1 \times L \subset \RR^4 \times S^1 \times X,
\label{surf_m5}
\end{equation}

\noindent where $L \cong S^1 \times \CC$ is a Lagrangian submanifold in $X$ \cite{harvey.lawson} such that 
an end-point of $L$ is on an edge of the toric web diagram \cite{aganagic.vafa}. The $M5$-branes geometrically 
engineer simple-type half-BPS surface operators that reduce the gauge group to $SU(N-1) \times U(1)$ on the 
surface $\RR^{\, 2}$ \cite{gukov.witten, alday.tachikawa, braverman.feigin.finkelberg.rybnikov, kanno.tachikawa}.

\subsection{From surface operators to primary-field vertex operators}
The AGT/W correspondence \cite{alday.gaiotto.tachikawa, wyllard} relates a class of 4D $\cN = 2$ 
supersymmetric gauge theories on $\RR^4$ to 2D Toda conformal field theories. Each of these Toda conformal 
field theories is defined on a punctured Riemann surface that is related to the Seiberg-Witten curve of 
the gauge theory and to the mirror curve of the Calabi-Yau 3-fold $X$. 
The simple-type surface operators on the gauge theory side correspond to vertex operators that, in turn,
correspond to the highest-weight states in irreducible fully-degenerate highest-weight representations on 
the conformal field theory side \cite{alday.gaiotto.gukov.tachikawa.verlinde,dimofte.gukov.hollands}. 
In other words, the $M5$-branes in \eqref{surf_m5} 
correspond to primary-field vertex operators of fully-degenerate representations in Toda conformal field 
theory \cite{kozcaz.pasquetti.wyllard, dimofte.gukov.hollands, taki.02, awata.fuji.kanno.manabe.yamada}. 
From that it follows that a condensation of the $M5$-branes corresponds to a condensation of vertex operators.
We expect that such a condensation leads to an off-critical deformation of the chiral blocks in the conformal 
field theory of the type that leads to correlation functions in off-critical integrable models. We will say 
more about this in section \textbf{\ref{section.8}}.
In the following we show that for $X=\CC^3$, $M5$-brane condensates lead to 
the refined $qt$-MacMahon generating function \eqref{def_ref_macmahon}.

\subsection{A $qt$-partition function from two brane condensates}
\label{subsec.off_macmahon}

\subsubsection{The normalized version of the computation}

\begin{figure}[t]
\centering
\includegraphics[width=60mm]{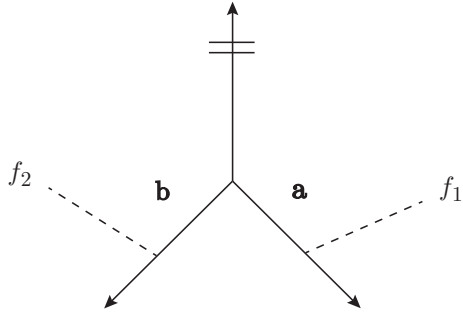}
\caption{The vertex with two brane stacks with open-string moduli 
$\mathbf{a}=\ll a_I \rr_{I=1,2,\ldots}$, 
$\mathbf{b}=\ll b_I \rr_{I=1,2,\ldots}$ and framing factors $f_1$ and $f_2$.}
\label{2-leg}
\end{figure}

Starting from the refined open-string partition function on $\CC^3$, which is the 
refined vertex, we trivialize the Young diagram on one of the three legs, and add 
a stack of infinitely-many branes on each of the two other legs. The first stack 
has open-string moduli $\mathbf{a} = \ll a_1, a_2, \ldots \rr$, and framing factor 
$f_1$, and the second has $\mathbf{b} = \ll b_1, b_2, \ldots \rr$, and framing factor 
$f_2$, as indicated in \figref{2-leg}. 
The result is the open-string partition function\footnote{\,
We take the holonomies along the un-preferred legs to be proportional 
to Schur functions \cite{Kozcaz:2018ndf} 
(see also \cite{kozcaz.pasquetti.wyllard, dimofte.gukov.hollands, Iqbal:2011kq}).
In the absence of the condensates, we have a closed string partition function on 
$\CC^3$. The M5-branes that condense are equivalent to open strings.
},

\begin{align}
Z_{\, x \, y}^{\, \ll f_1, \, f_2 \rr}\ll \mathbf{a}, \mathbf{b} \rr =
\cN_{\, \textit{branes}} \, \sum_{Y_{\, 1},Y_{\, 2}} 
\ll \prod_{i=1,2} f_{\, Y_i}\ll x, y \rr^{f_i} \rr
\cR_{\, Y_{\, 1}\, Y_{\, 2}\, \emptyset}\ll x, y \rr
s_{\,Y_{\, 1}}\ll \mathbf{a} \rr s_{\,Y_{\, 2}}\ll \mathbf{b} \rr,
%= 
%\\
%\cN_{\, \textit{branes}} \, \sum_{Y_{\, 1},Y_{\, 2}, Y}
%x^{\, \frac12 \, \ll \ll f_1+1 \rr \, \kappa_{Y_{\, 1}} + f_2 \, \kappa_{Y_{\, 2}} \rr} \,
%s_{\,Y_{\, 1}/Y}\ll x^{-\rho} \rr s_{\,Y_{\, 2}^{\, \prime}/Y}\ll x^{-\rho} \rr
%s_{\,Y_{\, 1}}\ll \ll -1 \rr^{f_1+1} \mathbf{a} \rr s_{\,Y_{\, 2}}\ll \ll -1 \rr^{f_2+1} \mathbf{b} \rr,
\label{computation.01}
\end{align}

\noindent where $\cN_{\, \textit{branes}}$ is a normalization factor, due to the introduction of the branes 
to be determined in the sequel. 
Choosing $\ll f_1, f_2 \rr = \ll -1, 0 \rr$ we get
\begin{align}
Z_{\, x \, y}^{\, \ll -1, \, 0 \rr}\ll \mathbf{a}, \mathbf{b} \rr =
\cN_{\, \textit{branes}} \,
\sum_{Y_{\, 1},Y_{\, 2},Y}
v^{\, -|Y|}\,
s_{\, Y_{\, 1} / Y} \ll y^{\, - \rho} \rr 
s_{\,Y_{\, 1}} \ll v \mathbf{a} \rr
s_{\, Y_{\, 2}^{\, \prime} / Y} \ll x^{\, - \rho} \rr
s_{\,Y_{\, 2}} \ll -v^{-1} \mathbf{b} \rr,
\end{align}

\noindent where $v = \ll x /y \rr^{1/2}$.
Using the Cauchy identities in appendix \textbf{A} we obtain,

\begin{equation}
Z_{\, x \, y}^{\, \ll -1, \, 0 \rr} \ll \mathbf{a}, \mathbf{b} \rr =
\cN_{\, \textit{branes}} \,
\ll \prod_{I,J=1}^{\infty} \ll 1 - v^{-1} a_I b_J \rr \rr
\ll \prod_{I=1}^{\infty}\frac{L\ll v^{-1} \, b_I, x\rr}{L\ll v \, a_I, y\rr} \rr,
\label{2_legged}
\end{equation}

\noindent where $L \ll a, x\rr$ is the quantum dilogarithm,  

\begin{equation}
L \ll a, x\rr = \prod_{m \, = \, 1}^{\infty} \ll 1 - a \, x^{\, m - \frac12} \rr =
\exp{\ll \sum_{n \, = \, 1}^{\infty} \frac{a^n}{n \ll x^{\, n/2} - x^{\, - \, n/2} \rr}\rr}
\label{q_dilog}
\end{equation}

\noindent The partition function \eqref{2_legged} includes 
the contribution of the brane-brane interactions across the brane-stacks. 
To remove this contribution, we take the normalization factor 
$\cN_{\, \textit{branes}}$ to be, 

\begin{equation}
\cN_{\, \textit{branes}} = 
\prod_{I, J = 1}^{\infty} 
\frac{1}{\ll 1 - v^{-1} a_I b_J \rr}, 
\label{norm_fixed}
\end{equation}

\noindent and obtain the partition function without the brane-brane 
interactions,

\begin{equation}
Z_{\, x \, y}^{\, \ll -1, \, 0 \rr} \ll \mathbf{a}, \mathbf{b} \rr =
\prod_{I=1}^{\infty}\frac{L\ll v^{-1} b_I, x\rr}{L\ll v a_I, y\rr},\quad 
v=\ll \frac{x}{y} \rr^{1/2}
\label{2_legged.normalized}
\end{equation}

\subsubsection{The unnormalized version of the computation}
The above calculation started from the normalized vertex $\cR_{\, Y_1 \, Y_2 \, Y_3} \ll x, y \rr$. 
If we use the unnormalized vertex 
$\cR^{\, \prime}_{\, Y_1 \, Y_2 \, Y_3} \ll x, y \rr$ in \eqref{norm_ref_v}, 
we get the unnormalized partition function with a single-brane insertion and 
two condensates, 

\begin{align}
Z_{\, x \, y}^{\, \prime \, \ll -1, \, 0 \rr} \ll \mathbf{a}, \mathbf{b} \rr =
M_{\, x \, y} 
\prod_{I=1}^{\infty}\frac{L\ll v^{-1} b_I, x\rr}{L\ll v a_I, y\rr}
%= 
%\exp{\ll \sum_{n=1}^{\infty}\frac{1}{n \ll x^{\, n/2} - x^{\, - \, n/2} \rr^{\, 2}} 
%+
%\sum_{n = 1}^{\infty} 
%\sum_{I = 1}^{\infty} 
%\frac{
%\ll v^{-1} b_I\rr ^{\, n}
%}{
%n \ll x^{\, n/2} - x^{\, -\, n/2} \rr}
%-
%\sum_{n = 1}^{\infty} 
%\sum_{I = 1}^{\infty} 
%\frac{
%\ll v a_I\rr ^{\, n}
%}{
%n \ll y^{\, n/2} - y^{\, -\, n/2} \rr}\rr
%}
\label{off_closed}
\end{align}

\noindent Splitting the index $I \to (i,j)$, as in section \textbf{\ref{splitting.indices}}, 
and setting, 

\begin{equation}
a_I\, \to\, a_{ij}=q^{\, i}\, x^{j - \frac12},\quad 
b_I\, \to\, b_{ij}=q^{i-1}\, t\, x\, y^{j - \frac32}, \quad
i, j = 1, 2, \ldots,
\label{special_cl}
\end{equation}

\noindent we find, 

\begin{equation}
Z_{\, x \, y}^{\, \prime \, \ll -1, \, 0 \rr} \ll a_{ij}, b_{ij} \rr=
M_{\, x \, y} 
\prod_{i,j=1}^{\infty}\frac{L\ll v^{-1} b_{ij}, x\rr}{L\ll v a_{ij}, y\rr}=
\prod_{i \, = \, 0}^{\infty} \prod_{m,n=1}^{\infty} 
\frac{1-q^{\, i} t x^{\, m} y^{n-1}}{1- q^{\, i} x^{\, m} y^{n-1}}=
M_{\, x \, y}^{\, q \, t}
\label{relation.01}
\end{equation}

\noindent We conclude that the refined open-string partition function on $\CC^3$ with 
two condensates, with moduli as in \eqref{special_cl}, agrees with the refined 
$qt$-MacMahon generating function $M_{\, x \, y}^{\, q \, t}$ in \eqref{def_ref_macmahon} 
which gives the refined $qt$-deformed closed string partition function on $\CC^3$.
By taking the unrefined limit $y \to x$ in \eqref{relation.01}, we obtain,

\begin{equation}
Z_{\, x \, x}^{\, \prime \, \ll -1, \, 0 \rr} \ll a_{ij}, b_{ij} \rr=
M_{\, x \, x} 
\prod_{i,j=1}^{\infty}\frac{L\ll b_{ij}, x\rr}{L\ll a_{ij}, x\rr}=
\prod_{i \, = \, 0}^{\infty} \prod_{m \, = \, 1}^{\infty} 
\ll \frac{1-q^{\, i} t x^{\, m}}{1- q^{\, i} x^{\, m}} \rr^{\, m}=
M_{\, x \, x}^{\, q \, t},
\label{relation.02}
\end{equation}

\noindent where the right hand side is the $qt$-MacMahon generating function 
\eqref{vuletic_mac} of Vuleti\'c \cite{vuletic}, and the left hand side can 
be derived using the original vertex 
$\cC_{\, Y_1 \, Y_2 \, Y_3} \ll x\rr$ as in \cite{halmagyi.sinkovics.sulkowski}.

\subsubsection{Remark} The relations \eqref{relation.01} and \eqref{relation.02} 
agree with the result that conformal blocks computed using 
$\cM^{\, q \, t}_{\, Y_1 \, Y_2 \, Y_3} \ll x, y \rr$ are equal to those computed using 
$\cR_{\, Y_1 \, Y_2 \, Y_3} \ll x, y \rr$ up to a $qt$-dependent factor \cite{foda.wu.02}.

%%%%%%%%%%%%%%%%%%%%%%%%%%%%%%%%%%%%%%%%%%%%%%%%%%%%%%%%%%%%%%%
%%%%%%%%%%%%%%%%%%%%%%%%%%%%%%%%%%%%%%%%%%%%%%%%%%%%%%%%%%%%%%%

\section{A $qt$-partition function with a single-brane insertion from brane condensates}
\label{section.05}
%%%%%%%%%%%%%%%%%%%%%%%%%%%%%%%%%%%%%%%%%%%%%%%%%%%%%%%%%%%%%%%
%%%%%%%%%%%%%%%%%%%%%%%%%%%%%%%%%%%%%%%%%%%%%%%%%%%%%%%%%%%%%%%
\noindent \textit{We give an example of a refined $qt$-deformed partition function 
with a single-brane insertion that is obtained from its undeformed counterpart via 
brane condensation.}
\medskip 

\subsection{A partition function with two brane condensates and a single-brane insertion}

\begin{figure}[t]
\centering
\includegraphics[width=60mm]{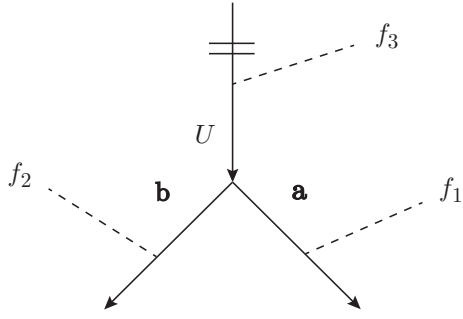}
\caption{The partition function with a single-brane insertion that has an open-string modulus $U$ 
and framing factor $f_3$, 
and two brane stacks that have open-string moduli 
$\mathbf{a}=\ll a_1, a_2, \ldots \rr$, 
$\mathbf{b}=\ll b_1, b_2, \ldots \rr$ and framing factors $\ll f_1, f_2 \rr$. 
By the inverse arrow in the preferred leg we assign the transpose $Y_{\, 3}^{\, \prime}$ 
of the Young diagram $Y_{\, 3}$, where we note the relation $\cR_{\, Y_{\, 1} \, Y_{\, 2} \, Y_{\, 3}} \ll x, x \rr = \cC_{\, Y_{\, 1} \, Y_{\, 2} \, Y_{\, 3}^{\, \prime}} \ll x \rr$ in \eqref{rel_CR}.}
\label{3-leg}
\end{figure}

Consider the same partition function as in section \ref{subsec.off_macmahon}, but now with 
an additional single brane (on the preferred leg of the refined vertex that has no brane-stacks)\footnote{\, 
Following \cite{kozcaz.pasquetti.wyllard, dimofte.gukov.hollands, Iqbal:2011kq, Kozcaz:2018ndf}, 
when a brane is inserted along the preferred leg of a refined topological vertex, we 
need to take the holonomy to be a Macdonald function rather than a Schur function. 
However, in the case of a single brane insertion, as discussed in the present work, 
a Macdonald function reduces to a Schur function, and we can take a Schur function 
as the holonomy.
} 
with an open-string modulus $U$ and a framing factor $f_3$, and two brane stacks, 
as represented in \figref{3-leg}, 

\begin{multline}
Z^{\, \ll f_1, \, f_2, \, f_3 \rr}_{\, x \, y} \ll U; \mathbf{a}, \mathbf{b} \rr =
\\ 
\cN_{\, \textit{branes}} \,
\sum_{Y_{\, 1},Y_{\, 2},Y_{\, 3}} 
\ll \prod_{i=1,2,3} f_{\, Y_i}\ll x, y \rr^{f_i} \rr
\cR_{\, Y_{\, 1}\, Y_{\, 2}\, Y_{\, 3}}\ll x, y \rr
s_{\,Y_{\, 1}}\ll \mathbf{a} \rr s_{\,Y_{\, 2}}\ll \mathbf{b} \rr s_{\,Y_{\, 3}^{\, \prime}}\ll U \rr
\end{multline}

\noindent Here $\cN_{\, \textit{branes}}$ is the normalization factor introduced in 
\eqref{computation.01} and determined in \eqref{norm_fixed}, and the Schur function 
$s_{\,Y_{\, 3}}\ll U \rr$ with a single variable $U$ is non-zero only for Young 
diagrams with a single row $y_{\, 1}=d$. 
Choosing the framing factors as $\ll f_1, f_2, f_3\rr = \ll -1, 0, f \rr$, we get,

\begin{multline}
Z^{\, \ll -1, \, 0, \, f \rr }_{\, x \, y} \ll U; \mathbf{a}, \mathbf{b} \rr =
\\ 
\cN_{\, \textit{branes}} \,
\sum_{Y_{\, 1},Y_{\, 2},Y_{\, 3},Y} 
f_{\, Y_{\, 3}^{\, \prime}}\ll x, y \rr^{f} 
x^{\, \frac12 \left\Vert Y_{\, 3} \right\Vert^{\, 2}}
\ll \prod_{\wsq \in Y_{\, 3}} \frac{1}{1 - x^{\, A^+_{\, \wsq}} y^{\, L_{\, \wsq}}} \rr
s_{\,Y_{\, 3}}\ll -U \rr \, \times
\\ 
v^{\,-|Y|}\,
s_{\, Y_{\, 1} / Y} \ll y^{\, - \rho} x^{\, - Y_{\, 3}} \rr 
s_{\,Y_{\, 1}} \ll v \mathbf{a} \rr
s_{\, Y_{\, 2}^{\, \prime} / Y} \ll x^{\, - \rho} y^{\, - Y_{\, 3}^{\, \prime}} \rr
s_{\,Y_{\, 2}} \ll -v^{-1} \mathbf{b} \rr,
\end{multline}

\noindent where $v = \ll x /y \rr^{1/2}$. Using the Cauchy identities in appendix \textbf{A}, 
we obtain,

\begin{multline}
Z_{\, x \, y}^{\, \ll -1, \, 0, \, f \rr}\ll U; \mathbf{a}, \mathbf{b} \rr =
\cN_{\, \textit{branes}} \,
\ll \prod_{I, J = 1}^{\infty} \ll 1 - v^{-1} a_I b_J \rr \rr
\ll \prod_{I = 1}^{\infty} \frac{L \ll v^{-1} b_I, x \rr}{L \ll v a_I, y\rr} \rr \, 
\times
\\ 
\sum_{d \, = \, 0}^{\infty}
\frac{
x^{\frac12 \ll 1-f \rr \, d^{\, 2}} \ll \ll -1 \rr^{f + 1} y^{\, f/2} \, U \rr^{\, d}
}{
\prod_{m \, = \, 1}^{\, d} \ll 1 - x^{\, m} \rr}
\ll \prod_{I = 1}^{\infty}
\frac{\ll 1 - a_I x^{\, \frac12} \rr}{\ll 1 - a_I x^{\, \frac12 - d} \rr}
\prod_{m \, = \, 1}^{\, d}
\frac{\ll 1 - b_I x^{\, m - 1} y^{\, - \, \frac12} \rr}{\ll 1 - b_I x^{\, m - 1} y^{\, \frac12} \rr} \rr,
\label{3_legged_ref}
\end{multline}

\noindent where $L\ll a, x\rr$ is the quantum dilogarithm in \eqref{q_dilog}.

\subsubsection{Normalization} Dividing the partition function with two condensates 
and a single-brane insertion by its counterpart that has no single-brane insertion 
(\ref{2_legged}), we obtain the normalized partition function, 

\begin{multline}
Z_{\, x \, y}^{\, \prime \, \ll -1, \, 0, \, f \rr}\ll U; \mathbf{a}, \mathbf{b} \rr =
\\ 
\sum_{d \, = \, 0}^{\infty}
\frac{
x^{\, \frac12 \, \ll 1 - f \rr {d^{\, 2}}} \ll \ll -1 \rr^{\, f + 1} \, y^{\, f/2} \, U \rr^{\, d}
}{
\prod_{m \, = \, 1}^{\, d} \ll 1 - x^{\, m} \rr
}
\ll \prod_{I=1}^{\infty}
\frac{\ll 1 - a_I \, x^{\frac12} \rr}{\ll 1 - a_I \, x^{\frac12 - d} \rr}
\prod_{m \, = \, 1}^{\, d}
\frac{\ll 1 - b_I \, x^{\, m - 1} \, y^{-\frac12} \rr}{\ll 1 - b_I \, x^{\, m - 1} \, y^{\, \frac12} \rr} \rr
\label{off_brane}
\end{multline}

\noindent We now show that for a suitable choice of the moduli $\ll a_1, a_2, \ldots \rr$ and 
$\ll b_1, b_2, \ldots \rr$, the normalized partition function (\ref{off_brane}) of a single-brane 
insertion and two condensates is the $qt$-deformation of the partition function on $\CC^{\, 3}$ 
with a single-brane insertion (and no condensates). The latter without the $qt$-deformation 
is obtained from \eqref{off_brane} by setting the open-string moduli of the condensates to zero, 

\begin{equation}
Z_{\, x \, y}^{\, \prime \, \ll -1, \, 0, \, f \rr} \ll U; 0, 0 \rr =
\\
\sum_{d \, = \, 0}^{\infty}
\frac{
x^{\, \frac12 \, \ll 1-f \rr {d^{\, 2}}} \ll \ll -1 \rr^{f+1} \, y^{\, f/2} \, U \rr^{\, d}
}{
\prod_{m \, = \, 1}^{\, d} \ll 1 - x^{\, m} \rr
}
\label{br_pf1}
\end{equation}

\subsubsection{Remark} Using the specialization of the one-row Schur function,

\begin{equation}
s_{\, \ll d \rr}\ll x^{\,\rho} \rr =
x^{\, \frac12 \, d\ll d-1\rr}\, s_{\, \ll 1^{\, d} \rr} \ll x^{\,\rho} \rr =
\frac{ 
\ll -1 \rr^{\, d} x^{\, d^{\, 2}/2}
}{
\prod_{m \, = \, 1}^{\, d} \ll 1 - x^{\, m} \rr},
\end{equation}

\noindent the partition function \eqref{br_pf1} is expressed in terms of Schur 
functions as, 

\begin{align}
\begin{split}
Z_{\, x \, y}^{\, \prime \, \ll -1, \, 0, \, f \rr} \ll U; 0, 0 \rr &=
\sum_{Y} x^{\, - \frac12\, f\, d^{\, 2}} s_{\,Y}\ll x^{\,\rho} \rr 
s_{\,Y}\ll \ll -1 \rr^f \, y^{\, \frac12\, f} \, U \rr 
\\
&=
\sum_{Y} x^{\, \frac12 \ll 1-f \rr d^{\, 2}}s_{\,Y^{\, \prime}}\ll x^{\,\rho} \rr
s_{\,Y}\ll \ll -1 \rr^f \,  x^{\, -\frac12} \, y^{\, \frac12\, f} \, U \rr 
\end{split}
\end{align}

\noindent Using the Cauchy identities in appendix \textbf{A}, the special cases of 
\eqref{br_pf1} that correspond to $f = 0, 1$, satisfy, 

\begin{equation}
Z_{\, x \, y}^{\, \prime \, \ll -1, \, 0,  \, 0 \rr} \ll U; \, 0, \, 0 \rr = 
\ll Z_{\, x \, y}^{\, \prime \, \ll -1, \, 0, \, 1 \rr} \ll v\,U; \, 0, \, 0 \rr \rr^{-1} =
L \ll U, x \rr
\label{br_pf_qd}
\end{equation}

\subsection{The $qt$-deformation of 
$Z_{\, x \, y}^{\, \prime \, \ll -1, \, 0, \, 1 \rr} \ll U; \, 0, \, 0 \rr$} 
The $qt$-partition function on $\CC^{\, 3}$ with a single-brane insertion with an open-string 
modulus $U$, can be computed using the Macdonald vertex as,

\begin{equation}
Z_{\, x \, y}^{\, q \, t} \ll U \rr  = 
\sum_Y 
\cM_{\, \emptyset \, \emptyset \, Y}^{\,q\,t} \ll x, y \rr s_{\, Y^{\, \prime}} \ll U \rr =
\sum_{d \, = \, 0}^{\infty}
\ll \prod_{i \, = \, 0}^{\infty} \prod_{m \, = \, 1}^{\, d}
\frac{1 - q^{\, i} \, t \, x^{\, m}}{1 - q^{\, i} \, x^{\, m}} \rr U^{\, d}
\label{Z_brane_c3_qt}
\end{equation}

\noindent Using 
$Z_{\, x \, y}^{\, \prime \, \ll -1, \, 0, \, 1 \rr} 
\ll y^{\,-\frac12} U; \, 0, \, 0 \rr=Z_{\, x \, y}^{\,q\,q} \ll U \rr$, 
this $qt$-partition function can be considered as the $qt$-deformation 
of the undeformed partition function \eqref{br_pf1}.

\subsection{Identification}
To identify the partition function in \eqref{off_brane} with that in \eqref{Z_brane_c3_qt}, we make 
the choice of moduli, 

\begin{equation}
a_I \, \to \, a^{\, \prime}_{ij} = x^{\, d}\, a_{ij} = q^{\, i    }\, x^{j - \frac12 + d}, \quad 
b_I \, \to \, b^{\, \prime}_{ij} = y    \,    b_{ij} = q^{\, i - 1}\,  t\,  x\, y^{j - \frac12    }, \quad
i, j = 1, 2, \ldots
\label{special_op}
\end{equation}

\noindent instead of that in \eqref{special_cl}. In other words, in this case, the moduli 
of the condensates now depend on the length of the single-row Young diagram that labels 
the Schur function that characterizes the single-brane insertion, $d$. For this modified 
choice of moduli, the normalized partition function \eqref{off_brane} with a single-brane 
insertion and two condensates becomes, 

\begin{equation}
Z_{\, x \, y}^{\, \prime \, \ll -1, \, 0, \, f \rr} 
\ll y^{\, - \, \frac12 \, f } U; \, a^{\, \prime}_{i j}, \, b^{\, \prime}_{i j} \rr =
\sum_{d \, = \, 0}^{\, \infty}
x^{\, \frac12 \, \ll 1 - f \rr {d^{\, 2}}}
\ll \prod_{i \, = \, 0}^{\infty} \prod_{m \, = \, 1}^{\, d}
\frac{1 - q^{\, i} \, t \, x^{\, m}}{1 - q^{\, i} \, x^{\, m}} \rr \ll \ll -1 \rr^{f+1} U \rr^{\, d}, 
\label{off_brane_qt}
\end{equation}

\noindent and we find,

\begin{equation}
Z_{\, x \, y}^{\, \prime \, \ll -1, \, 0, \, 1 \rr} 
\ll y^{\, - \, \frac12} U; \, a^{\, \prime}_{i j}, \, b^{\, \prime}_{i j} \rr =
\sum_{d \, = \, 0}^{\infty}
\ll \prod_{i \, = \, 0}^{\infty} \prod_{m \, = \, 1}^{\, d}
\frac{1 - q^{\, i} \, t \, x^{\, m}}{1 - q^{\, i} \, x^{\, m}} \rr U^{\, d}=
Z_{\, x \, y}^{\, q \, t} \ll U \rr
\end{equation}

\noindent We conclude that the refined $qt$-partition function with a single-brane insertion 
(and no condensates) coincides with its undeformed counterpart (with condensates) for 
a suitable choice of the framing factors, and of the open-string moduli of the condensates. 
Note that this refined $qt$-partition function does not depend on $y$, and coincides with 
the result computed by the original vertex $\cC_{\, Y_{\, 1}\, Y_{\, 2}\, Y_{\, 3}}\ll x \rr$ 
in a similar way.

\subsubsection{Remark} We interpret the change in the choice of the moduli of the condensates 
from that in \eqref{special_cl} to that in \eqref{special_op} as a back-reaction of the condensates 
to the single-brane insertion. 

\subsubsection{Remark} 
We have shown that the $qt$-deformed partition functions \eqref{def_ref_macmahon} and 
\eqref{Z_brane_c3_qt} are obtained, in the absence of a $qt$-deformation,  
from the partition functions \eqref{off_closed} and \eqref{off_brane}, respectively. 
These results depend on the chosen specializations \eqref{special_cl} and \eqref{special_op} 
that were made to obtain results that can be clearly interpreted. 
A study of the special significance (if any) of the choices that were made and the consequences 
of more general choices is beyond the scope of the present work.

%%%%%%%%%%%%%%%%%%%%%%%%%%%%%%%%%%%%%%%%%%%%%%%%%%%%%%%%%%%%%%%
%%%%%%%%%%%%%%%%%%%%%%%%%%%%%%%%%%%%%%%%%%%%%%%%%%%%%%%%%%%%%%%

\section{$qt$-Deformations as geometric transitions}
\label{section.06}
\label{subsec_geom_trans}
%%%%%%%%%%%%%%%%%%%%%%%%%%%%%%%%%%%%%%%%%%%%%%%%%%%%%%%%%%%%%%%
%%%%%%%%%%%%%%%%%%%%%%%%%%%%%%%%%%%%%%%%%%%%%%%%%%%%%%%%%%%%%%%
\noindent \textit{We discuss the relation of the brane condensates and geometric transitions
in the context of unrefined objects.} 
\medskip 

\subsection{Brane condensates and geometric transitions}
Following Gomis and Okuda \cite{gomis.okuda.01, gomis.okuda.02}, brane insertions change the topology 
of a Calabi-Yau 3-fold \textit{via} a geometric transition \cite{gopakumar.vafa.01}, and a Calabi-Yau 
3-fold with brane insertions is equivalent to a bubbling Calabi-Yau 3-fold of a more complicated 
topology, but without brane insertions. Correspondingly, an interpretation of the result in section
\textbf{\ref{subsec.off_macmahon}} is that a condensate (which is a set of infinitely-many brane insertions) 
changes the topology of $\CC^3$ \textit{via} a geometric transition, and $\CC^3$ with condensates 
is equivalent to another Calabi-Yau 3-fold of a more complicated geometry, but without condensates.  
To test this interpretation, we consider the $qt$-MacMahon generating function $M^{\, q \, t}_{\, x \, x}$ 
in \eqref{vuletic_mac}, which, as we showed in section \textbf{\ref{subsec.off_macmahon}}, is equal to 
the open-string partition on $\CC^3$ with two condensates, and interpret it as an undeformed (no
condensates) closed string partition function on a Calabi-Yau 3-fold with more complicated topology 
than $\CC^3$. 

\subsection{Gopakumar-Vafa invariants}
The partition function $Z_X \ll x, \, \mathbf{Q} \rr$ of the string on a Calabi-Yau 3-fold $X$ with
(exponentiated) K\"ahler moduli $\mathbf{Q}$, is the generating function of Gopakumar-Vafa invariants 
$n_{\, \beta, \, g} \in \ZZ$ \cite{gopakumar.vafa.02},

\begin{equation}
Z_X \ll x, \, \mathbf{Q} \rr =
\exp \ll 
\sum_{\beta \, \in \, H_2 \ll X, \, \ZZ \rr} \,  
\sum_{g \, = \, 0}^{\infty} \, 
\sum_{n \, = \, 1}^{\infty} \, 
\frac{n_{\, \beta, \, g}}{n} \, \ll x^{\, n/2} - x^{\, - \, n/2} \rr^{2 \, g \, - \, 2}
\mathbf{Q}^{\, \beta \, n}
\rr,
\label{gopakumar.vafa}
\end{equation}

\noindent where we have followed the notation used in \cite{marino.review}. Namely, 
if $i = \ll 1, 2, \ldots, b_2 \rr$, where $b_2$ is the second Betti number of $X$,  
$S_i$ is a basis of the second homology group $H_2 \ll X, \, \ZZ \rr$,
and $Q_i$ are (exponentiated) K\"ahler parameters,
then for any $\beta = \sum_i n_i [S_i] \in \, H_2 \ll X, \, \ZZ \rr$, $n_i \in \ZZ$, 
$\mathbf{Q}^{\, \beta} = \prod_i Q_{\, i}^{\, n_i}$. 
Comparing $M^{\, q \, t}_{\, x \, x}$ in \eqref{vuletic_mac} normalized by $M_{\, x \, x}$ in \eqref{macmahon} 
and the expansion in \eqref{gopakumar.vafa}, 
we find that $n_{\, \beta, \, 0} = \pm \, 1$, $n_{\, \beta, \, g} = 0$, for $g = 1, 2, \ldots$, 
which are the Gopakumar-Vafa invariants of a genus-0 manifold with infinitely-many homology 
2-cycles $\beta$. 
From \eqref{special_cl}, the infinitely-many branes (in the unrefined case) have holonomies,

\begin{equation}
\log a_{ij}= g_s \, \ll i\, N_q - j + \frac12 \rr,
\quad
\log b_{ij}= g_s \, \ll (i-1)\, N_q + N_t - j + \frac12 \rr,
\end{equation}

\noindent where $g_s = - \log x$, $g_s N_q = \log q$, $g_s N_t = \log t$, and 
according to \cite{gomis.okuda.01, gomis.okuda.02}, after large $N_q$ and $N_t$ limit, 
this yields a Calabi-Yau 3-fold via the bubbling.
This agrees with our interpretation of the $qt$-deformation in terms of 
a geometric transition driven by a condensate, that is, the insertion of infinitely-many 
branes. In section \textbf{\ref{subsec.qt_qc}}, we identify this geometry with that of 
an infinite strip, but before we do that, we consider a simple, but important example. 

\subsection{A simple example of a geometric transition}\label{sub:simple_geom}
In the special case of $q = 0$, $t \ne 0$, the $qt$-MacMahon generating function \eqref{vuletic_mac} is, 

\begin{equation}
M^{\, 0 \, t}_{\, x \, x}
=
M_{\, x \, x} \, 
\prod_{m \, = \, 1}^{\infty} \ll 1 - t \, x^{\, m} \, \rr^{\, m}
\label{qt_def_cl_0t}
\end{equation}

\begin{figure}[t]
\centering
\includegraphics[width=130mm]{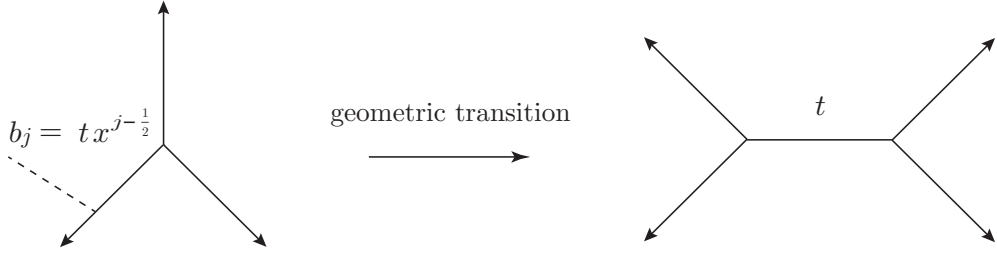}
\caption{The figure on the left represents the partition function with a single-brane insertion 
on $\CC^{\, 3}$. 
The figure on the right represents the closed string partition function on the resolved conifold.
They are related by a geometric transition.}
\label{geom_trans_res}
\end{figure}

\noindent This coincides with the undeformed closed string partition function on the resolved conifold, 
which is the total space of $\mathcal{O}(-1) \oplus \mathcal{O}(-1) \to \mathbb{P}^1$ with 
a single (exponentiated) K\"ahler modulus $t$, in agreement with the interpretation of the 
$t$-deformation of the MacMahon's generating function proposed in \cite{sulkowski}.\footnote{\,
What we call a $t$-deformation is called a $Q$-deformation in \cite{sulkowski}.} 
From the perspective of this section, what we have is the simple geometric transition in 
\figref{geom_trans_res}.

%%%%%%%%%%%%%%%%%%%%%%%%%%%%%%%%%%%%%%%%%%%%%%%%%%%%%%%%%%%%%%%
%%%%%%%%%%%%%%%%%%%%%%%%%%%%%%%%%%%%%%%%%%%%%%%%%%%%%%%%%%%%%%%

\section{$qt$-Quantum curves}
\label{section.07}
\label{subsec.qt_qc}
%%%%%%%%%%%%%%%%%%%%%%%%%%%%%%%%%%%%%%%%%%%%%%%%%%%%%%%%%%%%%%%
%%%%%%%%%%%%%%%%%%%%%%%%%%%%%%%%%%%%%%%%%%%%%%%%%%%%%%%%%%%%%%%
\noindent \textit{We discuss the $qt$-quantum curves associated with the unrefined 
limit of the refined $qt$-deformed partition function with a single-brane insertion 
in section \textbf{\ref{section.05}}.}

\medskip 

\subsection{The quantum curve for $Z^{\,q\,t} \ll U \rr$}
\subsubsection{Two operators}  
In the following, we need the operators $\widehat{U}$ and $\widehat{V}$, where 
$\widehat{U}$ acts as multiplication by a variable $U$, and 
$\widehat{V}$ acts as,  

\begin{equation}
\widehat{V} := x^{\, U \frac{d}{d U}},
\end{equation}

\noindent and satisfy the $x$-Weyl relation,

\begin{equation}  
\widehat{V}\, \widehat{U} \, = \, x \, \widehat{U} \, \widehat{V} 
\end{equation}

\subsubsection{The quantum curve}
\noindent The operators $\widehat{U}$ and $\widehat{V}$ act on $Z^{\,q\,t} \ll U \rr$, 
the unrefined limit of the refined $qt$-partition function with a single-brane insertion 
\eqref{Z_brane_c3_qt}, as, 

\begin{equation}
\widehat{U}\, Z^{\,q\,t} \ll U \rr = U \, Z^{\,q\,t} \ll   U \rr,
\quad
\widehat{V}\, Z^{\,q\,t} \ll U \rr =      Z^{\,q\,t} \ll x U \rr
\label{operators.01}
\end{equation}

\noindent From \eqref{operators.01}, it follows that $Z^{\,q\,t} \ll U \rr$ satisfies 
the $x$-difference equation, 

\begin{equation}
\widehat{A}^{\, q\,t} \ll \widehat{U}, \widehat{V} \rr \, Z^{\,q\,t} \ll U \rr :=
\ll 
\prod_{i \, = \, 0}^{\infty}
\ll 1 - q^{\, i}     \widehat{V} \rr - \widehat{U} \prod_{i \, = \, 0}^{\infty}
\ll 1 - q^{\, i} t x \widehat{V} \rr 
\rr 
\, Z^{\,q\,t} \ll U \rr = 0,
\label{A_hat_c3}
\end{equation}

\noindent which is \textit{the quantum curve} related to $Z^{\,q\,t} \ll U \rr$. 
As discussed below, \eqref{A_hat_c3} is a $qt$-version of the quantum curve of $\CC^{\, 3}$ 
in string theory 
\cite{aganagic.dijkgraaf.klemm.marino.vafa, 
dijkgraaf.hollands.sulkowski.vafa, dijkgraaf.hollands.sulkowski, gukov.sulkowski.02}. 

\subsubsection{The classical limit of the quantum curve}
\begin{figure}[t]
\centering
\includegraphics[width=110mm]{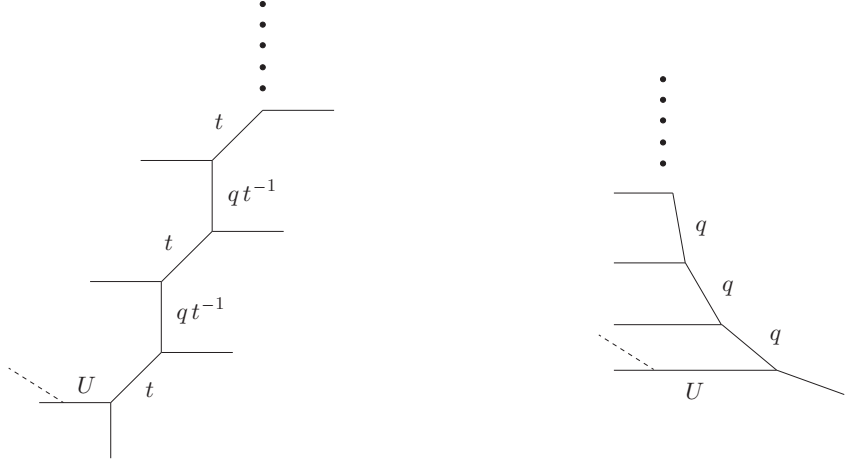}
\caption{
The figure on the left describes the infinite chain of $\ll -1, -1 \rr$ curves with K\"ahler moduli $t$ and $qt^{-1}$, 
and a brane insertion with the open-string modulus $U$.
The figure on the right describes the infinite chain of $\ll -2,  0 \rr$ curves with K\"ahler moduli $q$, 
and a brane insertion with the open-string modulus $U$.}
\label{strip_geom}
\end{figure}

Assuming that the asymptotic expansion of $Z^{\,q\,t} \ll U \rr $ in the classical limit, 
$g_s = - \log x \to 0$, has the WKB-form,  

\begin{equation}
Z^{\,q\,t}\ll U \rr  \sim 
\exp{
\ll 
- \, \frac{1}{g_s} \, \int^{\, U} \log V \ll U^{\, \prime} \rr \, 
\frac{
dU^{\, \prime}
}{
 U^{\, \prime} 
} \rr
},
\end{equation}

\noindent then $V \ll U \rr$ is a solution of the equation,

\begin{equation}
A^{\, q \, t} \ll U, V \ll U \rr \rr := 
\prod_{i \, = \, 0}^{\infty} \ll 1 - q^{\, i}   V \ll U \rr \rr - U 
\prod_{i \, = \, 0}^{\infty} \ll 1 - q^{\, i} t V \ll U \rr \rr 
= 0, 
\label{classic_c3_qt}
\end{equation}

\noindent which is \textit{the classical curve} related to $Z^{\,q\,t} \ll U \rr$.
This curve can be identified with the mirror curve related to the infinite-strip 
geometry that consists of an infinite chain of $\ll -1, -1 \rr$-curves, see the 
figure on the left in \figref{strip_geom} \cite{iqbal.kashani.poor} (see also \cite{fuji.iwaki.manabe.satake}). 
This infinite-strip geometry agrees with the picture of condensates in sections 
\textbf{\ref{section.04}}, \textbf{\ref{section.05}} and \textbf{\ref{section.06}}.
In the remainder of this section, we consider a number of spacial cases of quantum curves. 

\subsection{Case 1}
Choosing $q = t$, the partition function with a single-brane insertion \eqref{Z_brane_c3_qt} 
reduces to the undeformed partition function 
$Z_{\, x \, x}^{\, \prime \, \ll -1, \, 0, \, 1 \rr} \ll x^{\,-\frac12} U; \, 0, \, 0 \rr$ 
with a single-brane insertion on $\CC^{\, 3}$ in \eqref{br_pf1}, 

\begin{equation}
Z^{\,q\,q} \ll U \rr =
\sum_{d \, = \, 0}^{\infty} 
\frac{1}{\prod_{m \, = \, 1}^{\, d} \ll 1 - x^{\, m} \rr}\, U^{\, d} = 
Z_{\, x \, x}^{\, \prime \, \ll -1, \, 0, \, 1 \rr} \ll x^{\,-\frac12} U; \, 0, \, 0 \rr 
= L \ll x^{-\frac12} U, x\rr^{-1},
\end{equation}

\noindent and we find the quantum curve, 

\begin{equation}
\ll 1 - \widehat{V} - \widehat{U} \rr \, Z^{\,q \,q} \ll U \rr = 0
\label{q_curve_c3_qq}
\end{equation}

\noindent The classical limit, $g_s \to 0$, of the quantum curve \eqref{q_curve_c3_qq} 
gives a mirror curve of $\CC^{\, 3}$ \cite{aganagic.vafa}, 

\begin{equation}
1 - U - V = 0
\label{c_curve_c3}
\end{equation}

\noindent In other words, the $qt$-quantum curve \eqref{A_hat_c3} is a $qt$-version 
of the quantum curve \eqref{q_curve_c3_qq}, and \eqref{classic_c3_qt} is a $qt$-version 
of the mirror curve \eqref{c_curve_c3}. 

\subsection{Case 2}
Choosing $q=0$ and $t\ne 0$, the partition function with a single-brane insertion 
\eqref{Z_brane_c3_qt} reduces to, 

\begin{equation}
Z^{\,0\,t} \ll U \rr  =
\sum_{d \, = \, 0}^{\infty} 
\ll \prod_{m \, = \, 1}^{\, d}
\frac{1 - t \, x^{\, m}}{1 - x^{\, m}}\rr U^{\, d},
\end{equation}

\noindent and we find the $t$-version of the quantum curve of $\CC^{\, 3}$, 

\begin{equation}
\ll 1 - \widehat{V} - \widehat{U} + t \, \widehat{V} \, \widehat{U} \rr \, Z^{\,0 \, t} \ll U \rr = 0
\label{q_curve_c3_0t}
\end{equation}

\noindent Note that $Z^{\,0\,t}\ll U \rr$ agrees with the undeformed partition function with 
a single-brane insertion, up 
to framing ambiguities, on the resolved conifold with the K\"ahler modulus $t$ \cite{sulkowski}, 
and the classical limit, $g_s \to 0$, of the quantum curve \eqref{q_curve_c3_0t} is the mirror 
curve of the resolved conifold \cite{aganagic.vafa}, 

\begin{equation}
1 - U - V + t \, U \, V = 0
\end{equation}

\noindent In other words, the $t$-deformation of $\CC^{\, 3}$ is the resolved conifold 
as discussed in section \textbf{\ref{sub:simple_geom}}.

\subsection{Case 3}
Choosing $q \ne 0$ and $t=0$, the partition function with a single-brane insertion 
\eqref{Z_brane_c3_qt} reduces to, 

\begin{equation}
Z^{\,q\,0} \ll U \rr = 
\sum_{d \, = \, 0}^{\infty} \frac{1}{\prod_{i \, = \, 0}^{\infty} \prod_{m \, =\, 1}^{\, d} \,  
\ll 1 - \, q^{\, i} \, x^{\, m} \rr}\, U^{\, d},
\end{equation}

\noindent and the $q$-version of the quantum curve of $\CC^{\, 3}$ is, 

\begin{equation}
\ll \prod_{i \, = \, 0}^{\infty} \ll 1 - q^{\, i} \, \widehat{V} \rr - \widehat{U} \rr \, 
Z^{\,q \, 0} \ll U \rr = 0
\label{q_curve_c3_q0}
\end{equation}

\noindent $Z^{\,q\,0}\ll U \rr$ agrees with the undeformed partition function with a single-brane
insertion, up to framing ambiguities and a slight modification of the K\"ahler moduli, for 
the infinite chain of $\ll -2, 0 \rr$-curves,

\begin{equation}
\mathcal{O} \ll -2 \rr \, \oplus \, \mathcal{O} \ll 0 \rr \to \mathbb{P}^1,  
\end{equation}

\noindent with the same K\"ahler modulus $q$ for all $\mathbb{P}^1$, see the figure on 
the right in \figref{strip_geom} \cite{iqbal.kashani.poor} (see also \cite{fuji.iwaki.manabe.satake}). 
This infinite-strip geometry can be obtained from that in the figure on the left in \figref{strip_geom} 
by suitable blow-downs.\footnote{\,
Starting from the infinite-strip geometry on the left in \figref{strip_geom}, 
one can think of what happens in the limit $t \to 0$ as follows. 
As $t \to 0$, the K\"ahler parameters $t$ vanish, while the K\"ahler parameters $q/t$ diverge, 
and the corresponding consecutive edges in the toric diagram combine in pairs to form a toric 
diagram that has edges with finite K\"ahler parameters $q$. 
The new infinite-strip geometry is on the right in \figref{strip_geom}.} 
The classical limit, $g_s \to 0$, of the quantum curve \eqref{q_curve_c3_q0} 
is the mirror curve of this strip geometry,  

\begin{equation}
\prod_{i \, = \, 0}^{\infty}\ll 1 - q^{\, i} \, V \rr \, - \, U = 0
\end{equation}

\noindent We conclude that the $q$-deformation of $\CC^{\, 3}$ is identified with the infinite-strip 
geometry in the figure on the right in \figref{strip_geom}, and that this infinite-strip geometry 
is the result of a geometric transition caused by the condensates.

%%%%%%%%%%%%%%%%%%%%%%%%%%%%%%%%%%%%%%%%%%%%%%%%%%%%%%%%%%%%%%%
%%%%%%%%%%%%%%%%%%%%%%%%%%%%%%%%%%%%%%%%%%%%%%%%%%%%%%%%%%%%%%%

\section{Remarks}
\label{section.8}
%%%%%%%%%%%%%%%%%%%%%%%%%%%%%%%%%%%%%%%%%%%%%%%%%%%%%%%%%%%%%%%
%%%%%%%%%%%%%%%%%%%%%%%%%%%%%%%%%%%%%%%%%%%%%%%%%%%%%%%%%%%%%%%
\noindent \textit{We collect a number of remarks, with particular attention to the interpretation 
of the various parameters that can appear in topological vertices, and to the relation with conformal 
field theory.} 
\medskip

\subsection{The AGT counterpart of brane condensates}
We showed that the Macdonald-type $qt$-deformation introduced in \cite{vuletic}, when applied to 
topological string partition functions \cite{foda.wu.02}, leads to $qt$-partition functions that 
are equivalent to partition functions without a $qt$-deformation but with condensates. 
These condensates are surface operator condensates, and their counterparts on the conformal 
field theory side of the AGT correspondence are vertex operator condensates in 2D chiral conformal 
blocks. 
While this has not been studied in any detail, we expect that these vertex operator condensates 
play, at the level of conformal blocks, the same role that switching-on off-critical perturbations 
plays, at the level of the correlation functions \cite{zamolodchikov}, and that results in correlation 
functions in 2D off-critical integrable models.
This expectation coincides with the results in 
\cite{awata.yamada.01, awata.yamada.02, nieri.pasquetti.passerini, 
nieri.pasquetti.passerini.torrielli, pasquetti}.\footnote{\,
See further discussion on section \textbf{\ref{discussion.01}}.}

\subsection{Four parameters}
\label{4.parameters}
If we start from a 4D instanton partition function in the absence of an $\Omega$-background, or an AGT-equivalent 
conformal block in a Gaussian 2D conformal field theory with an integral central charge, there are four known ways 
to modify such a partition function, or conformal block, and each of these ways is characterized by a parameter. 

\subsubsection{The radius of the $M$-theory circle, $R$}
\label{discussion.01}
Topological string partition functions are 5D objects, and the corresponding instanton partition functions 
live in $\RR^4 \times S^1$, where $S^1$ is the $M$-theory circle. 
For small $R$, one can think of the 5D instanton partition functions as $R$-deformations of their 4D limits, 
in the sense that switching on $R$ gradually is equivalent to including the lighter Kaluza-Klein modes that 
are infinitely-massive in the $R \rightarrow 0$, and that acquire finite masses as $R$ increases 
\cite{iqbal.kaplunovsky}.  
In 2D conformal field theory terms, switching $R$ on is equivalent to deforming the chiral conformal blocks 
away from criticality to obtain expectation values of type-I vertex operators 
\cite{davies.foda.jimbo.miwa.nakayashiki}, in some off-critical integrable statistical mechanical 
models \cite{awata.yamada.01, awata.yamada.02, nieri.pasquetti.passerini, 
nieri.pasquetti.passerini.torrielli, pasquetti}. 

\subsubsection{The refinement parameter $x/y$}
Starting with 4D instanton partition functions in the absence of an $\Omega$-background, one can 
switch on Nekrasov's $\Omega$-deformation parameters, that is $\eo + \et \neq 0$. In the presence 
of a finite $M$-theory circle of radius $R$, setting 
$x = e^{\, - \, R \, \eo}$, and 
$y = R^{\,      R \, \et}$, 
this refinement is equivalent to setting $x / y \neq 1$. In 2D conformal field theory terms, we 
modify the central charge of the conformal field theory while preserving conformal invariance, 
and the underlying statistical mechanical model remains critical. 

\subsubsection{The Macdonald deformation parameter $q/t$}
The $q/t$-deformation of \cite{vuletic, foda.wu.02} is yet another perturbation but, so far, 
no interpretation of this deformation is known. The purpose of this work is to offer one such 
interpretation. 

\subsubsection{The elliptic nome $p$}
In \cite{zhu, foda.zhu}, two versions were proposed of a topological vertex based on Saito's 
elliptic deformation of the quantum toroidal algebra $U_q \ll \widehat{\widehat{gl_1}} \rr$ 
\cite{saito.01, saito.02, saito.03}. 
In addition to the refinement 
parameters $\ll x, y\rr$, and the Macdonald-type deformation parameters $\ll q, t\rr$, this vertex 
depends on an elliptic nome parameter $p$ and copies of the $\ll q=t \rr$-limit of this vertex 
can be glued to obtain elliptic conformal blocks. The latter are equal to the elliptic conformal 
blocks that were computed in \cite{iqbal.kozcaz.yau, nieri} by gluing copies of the refined vertex 
of \cite{iqbal.kozcaz.vafa}, then gluing pairs of external legs. 

\subsection{Three off-critical deformations}
Aside from the refinement parameter $x/y$, which preserves criticality, it appears that we have 
three parameters that push the underlying 2D conformal blocks off-criticality, namely the $M$-theory 
circle radius $R$, the Macdonald parameter $q/t$, and the nome parameter $p$. 
One can show by explicit computation that these three parameters coexist and that their effects are 
different, but it remains unclear how to interpret these effects in statistical mechanics terms. 

\subsection{BPS states in $M$-theory}
Following \cite{gopakumar.vafa.02, Ooguri:1999bv}, topological string partition functions on 
a Calabi-Yau 3-fold encode the degeneracies of the BPS states in $M$-theory compactified on 
the Calabi-Yau 3-fold, and the interpretation of the $xy$-refinement (of the refined topological 
vertex) was discussed in \cite{iqbal.hollowood.vafa, Gukov:2004hz}.
What is the interpretation of the $qt$-deformation (of the Macdonald vertex) in the context 
of $M$-theory? 
In section \textbf{\ref{section.06}}, we argued that 
a topological string partition function 
on a Calabi-Yau 3-fold with finitely-many homology 2-cycles, 
\textit{in the presence of} a $qt$-deformation 
is equal, after a geometric transition, to a corresponding 
topological string partition function 
\textit{in the absence of} a $qt$-deformation, 
on a Calabi-Yau 3-fold with infinitely-many homology 2-cycles. 
From this correspondence, we expect that the $qt$-partition functions encode the degeneracies 
of BPS states in $M$-theory compactified on the Calabi-Yau 3-fold with infinitely-many homology 
2-cycles. 
A more direct and perhaps deeper interpretation at the level of the original Calabi-Yau 3-fold 
with finitely-many homology 2-cycles is beyond the scope of the present work.

\subsection{Summary}
In \cite{awata.kanno.01, awata.kanno.02, iqbal.kozcaz.vafa}, a refinement of the original topological 
vertex was obtained, and the physical meaning of this refinement was clear and related to switching-on 
a non-self-dual $\Omega$-background. In \cite{vuletic}, an independent Macdonald-type $qt$-deformation 
of MacMahon's generating function of plane partitions was obtained, and was used in \cite{foda.wu.02}
to $qt$-deformed the refined topological vertex, but no physical meaning of this deformation was proposed. 
In the present work, we have presented a number of simple but clear examples of $qt$-deformed topological 
string partition functions, and showed in sections \textbf{\ref{section.04}} and \textbf{\ref{section.05}} 
that, in these cases, the $qt$-deformation is equivalent to switching-on infinitely-many brane insertions, 
or equivalently brane condensates. 
In section \textbf{\ref{section.06}}, we showed that 
a Calabi-Yau 3-fold with a simple topology in the presence of these condensates is equivalent to 
another Calabi-Yau 3-fold with a more complicated topology without condensates, and argued that 
the condensates cause the Calabi-Yau 3-fold on which the topological string theory is formulated 
to undergo a geometric transition that changes its topology.
Finally, in section \textbf{\ref{section.07}}, we studied the $qt$-quantum curves related to the 
unrefined limit of the $qt$-partition functions studied in section \textbf{\ref{section.05}}, 
and showed that their classical limit does indeed correspond to undeformed partition functions 
on infinite-strip geometry, in agreement with the conclusion that the $qt$-deformation is equivalent 
to brane condensates that drive a geometric transition. We expect these conclusions to hold for 
$qt$-deformations of more complicated topological string partition functions.  

\appendix
%%%%%%%%%%%%%%%%%%%%%%%%%%%%%%%%%%%%%%%%%%%%%%%%%%%%%%%%%%%%%%%
%%%%%%%%%%%%%%%%%%%%%%%%%%%%%%%%%%%%%%%%%%%%%%%%%%%%%%%%%%%%%%%

\section{Useful Schur function identities}\label{app_formula}
\label{appendix}
%%%%%%%%%%%%%%%%%%%%%%%%%%%%%%%%%%%%%%%%%%%%%%%%%%%%%%%%%%%%%%%
%%%%%%%%%%%%%%%%%%%%%%%%%%%%%%%%%%%%%%%%%%%%%%%%%%%%%%%%%%%%%%%
\medskip

\noindent
The skew Schur functions satisfy the identities,

\begin{align}
s_{\,Y} \ll x^{\,\rho} \rr &= x^{\, \frac12 \kappa_{\, Y}} s_{\,Y^{\, \prime}} \ll x^{\,\rho} \rr,
\\
s_{\,Y}\ll x^{\,-\rho} \rr &=
x^{\, \frac12 \, \left\Vert Y^{\, \prime} \right\Vert^{\, 2}}
\prod_{\wsq \in Y} \frac{1}{1 - x^{\, H_{\, \wsq}}},
\\
s_{\,Y/\emptyset} \ll \mathbf{x} \rr &= s_{\,Y} \ll \mathbf{x} \rr,
\\
s_{\,Y_{\, 1}/Y_{\, 2}} \ll \mathbf{x} \rr &= 0 \quad\ \textrm{for}\ Y_{\, 1} \not\supset Y_{\, 2},
\\
s_{\,Y_{\, 1}/Y_{\, 2}} \ll c\,\mathbf{x} \rr &=
c^{\,\left| Y_{\, 1}\right| \, - \, \left| Y_{\, 2} \right|}\, s_{\,Y_{\, 1} / Y_{\, 2}} \ll \mathbf{x} \rr, \quad c\in \CC,
\\
s_{\,Y_{\, 1} / Y_{\, 2}} \ll x^{\, \rho \, + \, Y} \rr &= \ll -1 \rr^{\, \left| Y_{\, 1} \right| \, - \, \left| Y_{\, 2} \right|}\, 
s_{\,Y_{\, 1}^{\, \prime} / Y_{\, 2}^{\, \prime}} \ll x^{\, - \, \rho \, - \, Y} \rr
\label{schur_ids}
\end{align}

\noindent The Cauchy identities for the skew Schur functions are,

\begin{align}
\sum_Y  s_{\,Y/Y_{\, 1}} \ll \mathbf{x} \rr s_{\,Y/Y_{\, 2}} \ll \mathbf{y} \rr &=
\sum_Y  s_{\,Y_{\, 2}/Y} \ll \mathbf{x} \rr s_{\,Y_{\, 1}/Y} \ll \mathbf{y} \rr
\prod_{i,j=1} \ll \frac{1}{1 - x_i y_j} \rr,
\\
\sum_Y  s_{\,Y/Y_{\, 1}} \ll \mathbf{x} \rr s_{\,Y^{\, \prime}/Y_{\, 2}} \ll \mathbf{y} \rr &=
\sum_Y  s_{\,Y_{\, 2}^{\, \prime}/Y} \ll \mathbf{x} \rr s_{\,Y_{\, 1}^{\, \prime}/Y^{\, \prime}} \ll \mathbf{y} \rr
\prod_{i,j=1} \ll 1 + x_i y_j \rr
\label{cauchy_ids_sc}
\end{align}

\section*{Acknowledgements}
\noindent We thank Piotr Su{\l}kowski for useful discussions, the anonymous referee for questions 
and remarks that helped us improve the presentation, and the Mathematical Research Institute 
MATRIX, in Creswick, Victoria, Australia, for hospitality during the workshop \textit{\lq Integrability 
in Low-Dimensional Quantum Systems\rq}, where the present work was started. 
The work of OF is supported by the Australian Research Council Discovery Grant DP140103104.    
The work of MM was supported by the ERC Starting Grant no. 335739 \textit{\lq Quantum fields 
and knot homologies\rq} funded by the European Research Council under the European Union's 
Seventh Framework Programme, and currently by the Max-Planck-Institut f\"ur Mathematik in Bonn.

\end{document}